\documentclass[longauth]{aa}  
\usepackage{placeins}

\usepackage{graphicx}
\usepackage[flushleft]{threeparttable}

\usepackage{txfonts}
\usepackage[citecolor=blue, linkcolor=blue]{hyperref}
\hypersetup{colorlinks=True}
\newcommand{\Msun}{M$_{\odot}$}

\begin{document} 

\title{Physical properties of extreme emission-line galaxies at $z\sim4-9$ from the JWST CEERS survey}
 
\author{Llerena, M.\inst{1,2}\fnmsep\thanks{\email{mario.llerenaona@inaf.it}}
\and Amor\'in, R. \inst{3,2,4}
\and Pentericci, L.\inst{1}
\and Arrabal Haro, P. \inst{5}
\and Backhaus, B. E. \inst{6}
\and Bagley, M. B. \inst{7}
\and Calabr\`o A. \inst{1}
\and Cleri, N. J. \inst{8,9}
\and Davis, K. \inst{6}
\and Dickinson, M. \inst{4}
\and Finkelstein, S. L. \inst{7}
\and Gawiser, E. \inst{10}
\and Grogin, N. A. \inst{11}
\and Hathi, N. P. \inst{11}
\and Hirschmann, M. \inst{12}
\and Kartaltepe, J. S. \inst{13}
\and Koekemoer, A. M.  \inst{11}
\and McGrath E. J. \inst{14}
\and Mobasher, B. \inst{15}
\and Napolitano, L. \inst{1,16}
\and Papovich, C. \inst{8,9}
\and Pirzkal, N. \inst{17}
\and Trump, J. R. \inst{6}
\and Wilkins, S. M.  \inst{18,19}
\and Yung, L.~Y.~A. \inst{11} 
          }

   \institute{INAF - Osservatorio Astronomico di Roma, Via di Frascati 33, 00078, Monte Porzio Catone, Italy
   \and 
   Departamento de Astronomía, Universidad de La Serena, Av. Juan Cisternas 1200 Norte, La Serena, Chile
   \and Instituto de Astrof\'{i}sica de Andaluc\'{i}a (CSIC), Apartado 3004, 18080 Granada, Spain
   \and
    ARAID Foundation. Centro de Estudios de F\'{\i}sica del Cosmos de Arag\'{o}n (CEFCA), Unidad Asociada al CSIC, Plaza San Juan 1, E--44001 Teruel, Spain
    \and {NSF’s National Optical-Infrared Astronomy Research Laboratory, 950 N. Cherry Ave., Tucson, AZ 85719, USA}
    \and {Department of Physics, 196A Auditorium Road, Unit 3046, University of Connecticut, Storrs, CT 06269, USA}
    \and {Department of Astronomy, The University of Texas at Austin, Austin, TX, USA}
    \and {Department of Physics and Astronomy, Texas A\&M University, College Station, TX, 77843-4242 USA}
    \and{George P.\ and Cynthia Woods Mitchell Institute for Fundamental Physics and Astronomy, Texas A\&M University, College Station, TX, 77843-4242 USA}
     \and {Department of Physics and Astronomy, Rutgers University, Piscataway, NJ 08854, USA}
    \and{Space Telescope Science Institute, 3700 San Martin Drive, Baltimore, MD 21218, USA}
     \and{Institute of Physics, Laboratory of Galaxy Evolution, Ecole Polytechnique Fédérale de Lausanne (EPFL), Observatoire de Sauverny, 1290 Versoix, Switzerland}
      \and {Laboratory for Multiwavelength Astrophysics, School of Physics and Astronomy, Rochester Institute of Technology, 84 Lomb Memorial Drive, Rochester, NY 14623, USA}
       \and {Department of Physics and Astronomy, Colby College, Waterville, ME 04901, USA}
 \and {Department of Physics and Astronomy, University of California Riverside, 900 University Avenue, Riverside, CA 92521, USA }
 \and {Dipartimento di Fisica, Università di Roma Sapienza, Città Universitaria di Roma - Sapienza, Piazzale Aldo Moro, 2, 00185, Roma, Italy}
  \and {ESA/AURA Space Telescope Science Institute}
    \and{Astronomy Centre, University of Sussex, Falmer, Brighton BN1 9QH, UK}
 \and{Institute of Space Sciences and Astronomy, University of Malta, Msida MSD 2080, Malta}
    }

   \date{Received ; accepted }

%

%
%
%

%
%

%
%
%
%
%
%
%
%
%
%

 
  \abstract
   {
   Extreme emission line galaxies (EELGs) are typically characterized by high equivalent widths (EWs) which are driven by elevated specific star formation rates (sSFR) in low-mass galaxies with subsolar metallicities and little dust. Such extreme systems are exceedingly rare in the local universe, but the number density of EELGs increases with increasing redshift. Such starburst galaxies are currently presumed strongly to be the main drivers of hydrogen reionization over $5.5 < z < 15$, which serves to motivate many of the searches for high-$z$ EELGs. 
   }   
   {
   We aim to characterize the physical properties of a sample of $\sim$ 730 EELGs at $4\lesssim z<9$ photometrically selected from the CEERS survey using JWST/NIRCam. We validate our method and demonstrate the main physical properties of a subset of EELGs using NIRSpec spectra.
   }
   {We create synthetic NIRCam observations of EELGs using empirical templates based on $\sim$ 2000 local metal-poor starbursts to select EELGs based on color-color criteria. We study their properties based on SED fitting and flux excess from emission lines in the photometric filters. 
   }
   {
   Our sample of EELGs has a mean stellar mass of $10^{7.84}$\Msun with high sSFRs from SED fitting with a mean value of $10^{-7.03}$ yr$^{-1}$. We consider a delayed-$\tau$ model for the star formation history and find our sample of EELGs are young with a mean value of the time after the onset of star formation of 45Myr. We find that they have similar line ratios to local metal-poor starburst galaxies with high log([OIII]/H$\beta$)$\gtrsim$0.4-1 which indicates that star formation may be the dominant source of ionization in these galaxies. Based on the photometric fluxes and morphologies, we find an increase of EW([OIII]+H$\beta$) with sSFR and $\Sigma_{\rm SFR}$, and a decrease with age and stellar mass.  The sample of EELGs can reach $\Sigma_{\rm SFR}>$ 10 \Msun yr$^{-1}$ kpc$^{-2}$ which indicate they are strong candidates of LyC leakers. Another indirect indicator is the high values of O32$>$5 that can be reached for some galaxies in the sample. This indicates that they may have the conditions to facilitate the escape of ionizing photons. 
   }

   \keywords{Galaxies: starburst --
                Galaxies: high-redshift --
                Galaxies: evolution --
                Galaxies: formation --
                Galaxies: ISM
               }
\titlerunning{Extreme emission-line galaxies at $z\sim4-9$ from the JWST CEERS survey}
\authorrunning{Llerena, M. et al.}

\maketitle
%

\section{Introduction} \label{sec:intro}

Over the past decades, several studies have discovered a population of galaxies that undergo very strong star formation events. They are called extreme emission line galaxies (EELGs) and are typically characterized by high equivalent widths (EWs) of hundreds to thousands of Angstroms which are driven by elevated specific star formation rates
(sSFRs) up to 10-100 Gyr$^{-1}$ in galaxies with 
stellar masses $\lesssim 10^9$\Msun, with subsolar metallicities and little dust \citep[e.g.][]{vanderWel2011,Maseda2014,Amorin2015,Forrest2017}. These large emission line EWs are most commonly seen in the [OIII]$\lambda\lambda$4959,5007 (hereafter [OIII]) and H$\alpha$ lines in the rest-frame optical. The nebular emission lines are often driven by the ionizing photons produced in massive
and short-lived O and B stars (or active galactic nuclei (AGN)), whereas the
surrounding rest-optical continuum includes the contribution from
longer-lived and less massive stars \citep[][for a review]{Eldridge2022}.

Such extreme systems are rare in the local
Universe, but the number density of EELGs increases with increasing redshift \citep[e.g.][]{Smit2014} and they are expected to be common in the Epoch of Reionization (EoR). Recent results indicate that the population of galaxies in the EoR shows a mean EW([OIII]+H$\beta$)=780\r{A} \citep[e.g.][]{Endsley2022}. Examples of EELGs in the local universe are the so-called HII galaxies \citep{Terlevich1991} or blueberry galaxies \citep{Yang2017}, and Green Pea galaxies \citep{Cardamone2009, Amorin2010, Amorin2012, Izotov2011}.
Besides allowing us to study extreme conditions in galaxies in the early universe, such systems are also essential to understanding the process of reionization. Such starburst galaxies are currently presumed to be the main drivers of hydrogen reionization over $5.5 < z < 15$ \citep[e.g.][]{Robertson2015,Finkelstein2019,Yung2020b,Yung2020a}, which serves to motivate many of the searches for high-$z$ EELGs as well. To understand the properties of EELGs, large and representative samples of EELGs must be assembled.

The study of EELGs in the local Universe up to intermediate redshifts has been a significant field in recent years because they resemble the properties of galaxies in the EoR with high EWs of UV and optical lines \citep{Smit2015,Hutchison2019}. With the advent of the JWST era \citep{Gardner2006,Gardner2023}, the discoveries of young galaxies are rapidly moving deep into the EoR, where such systems can be studied now directly up to $z \sim 9$ with NIRCam and NIRSpec \citep[e.g.][]{Withers2023,Simmonds2023,Endsley2023,Davis2023}. The first step is to find and characterize such systems to conduct detailed spectroscopic studies.

Many of the searches for EELGs select samples using either narrow- \citep[e.g.][]{Sobral2013,IglesiasParamo2022,Lumbreras-Calle2022}, broad- \citep[e.g.][]{vanderWel2011,Onodera2020,Kojima2020,Chen2023,Davis2023} or medium-band photometry \citep[e.g.][]{Cohn2018,Withers2023,Simmonds2023}, and slitless spectroscopy \citep[e.g.][]{Maseda2018,Kashino2022}. Using photometry, EELGs are selected based on the enhancing effect the emission line fluxes have in some specific filter, depending on its redshift. Unlike the commonly used Lyman break technique, broad-band color selection does not require high S/N in the rest-frame UV-optical continuum. Faint continuum EELGs are difficult to detect through their continuum emission alone but can be selected via their color excess due to the presence of emission lines. This population of very faint-continuum galaxies may play an essential role in reionization if found in great numbers \citep[e.g.,][]{Endsley2022}, yet will not be detected with many of the commonly used selection criteria.

Deep imaging with new JWST capabilities opens a new window to identify such young metal-poor starbursts. Our sample originates from the Cosmic Evolution Early Release Science survey \citep[CEERS,][]{Finkelstein2023,Finkelstein2023b} which is a JWST Early Release Science program that obtained imaging and spectroscopy of the Extended Groth Strip \citep[EGS, 14h19m00s +52$^{\circ}$48'00'', ][]{Davis2007} field with three instruments and five coordinated parallel observing modes. We use the CEERS NIRCam and NIRSpec data to look at EELGs over a redshift range $4 \lesssim z < 9$
to study their physical properties in detail to determine what physical conditions are necessary for a galaxy to be an EELG. We focus on the
[OIII]$\lambda$5007 line, but we also consider cases where H$\alpha$ may also have extreme EWs.

This paper is organized as follows: in Sec. \ref{sec:methods-jwst}, we present the observational data and the templates of EELGs we considered in this work. Particularly in Sec. \ref{sec:jwst:synt_nircam}, we present our NIRCam synthetic observations to estimate the color excess to identify EELG candidates. In Sec. \ref{sec:jwst_colordiagrams}, we present the color-color diagrams used and the final sample of EELG candidates. Regarding the analysis of the data, in Sec. \ref{sec:photo-fluxes} we present the photometric fluxes and EWs of emission lines. In Sec. \ref{sec:SED-nircam} we present the physical properties of our sample based on SED fitting and in Sec. \ref{sec:sizes} we present the physical sizes and the star formation rate (SFR) surface density of the galaxies in our sample. In Sec. \ref{sec:jwst_nirspec} we present the NIRSpec spectroscopy of a subsample of 47 galaxies in our candidate list. In Sec. \ref{sec:results-discussion}, we present our results and discussions on the ionization properties of the EELG candidates and the comparison with the control sample of non-EELG galaxies. Finally, in Sec. \ref{sec:jwst-summary}, we present our conclusions. 

Throughout this paper, we adopt a $\Lambda$-dominated flat universe with $\Omega_\Lambda = 0.7$, $\Omega_M = 0.3,$ and H$_0= 70$ km s$^{-1}$ Mpc$^{-1}$. All magnitudes are quoted in the AB system. Equivalent widths are quoted in the rest frame and are positive for emission lines. We consider log(O/H)$_{\odot}=8.69$ \citep{Asplund2009}.

\section{Data and sample selection}\label{sec:methods-jwst}

\subsection{The CEERS survey}
The complete CEERS program involves imaging with the NIRCam short and long-wavelength channels in ten pointings, observed as coordinated parallels to primary observations with the NIRSpec and the MidInfrared Instrument \citep[MIRI;][]{Wright2015}. In this paper, we used photometric data from NIRCam and NIRSpec spectra. The CEERS observations were split into two epochs. The first epoch was executed in June 2022, and the second in December 2022. An additional third epoch of NIRSpec prism spectra (pointings 11 and 12) was taken in February 2023, due to the MSA short that affected prism 9 + 10 in December 2022. In total, 10 pointings were taken with NIRCam imaging, including  seven filters per pointing (F115W, F150W, F200W, F277W, F356W, F410M, and F444W, see Fig. \ref{fig:wave-redshift}), reaching 5$\sigma$ depths of $\sim$28.5 - 29.2 AB mag for point sources for a total area
of $\sim$ 97 arcmin$^2$.

\begin{table*}[!t]
\caption{Rest-frame EWs of bright optical emission lines for the ASK templates considered EELGs. 
The obtained values are comparable with the measurements reported in \cite{SA2010}.}
    \label{tab:ask_classes}
    \begin{tabular}{|c|c|c|c|c|}\hline
    ASK class &EW(H$\beta$)&EW([OIII]$\lambda$5007)&EW(H$\alpha$)&log(O/H)+12$^*$\\
    &\r{A}&\r{A}&\r{A}&\\\hline
15 &169.6 &1097.9& 918.3&{8.00}$^{(a)}$\\ \hline
17& 144.9 &874.3& 782.3&{8.09}\\ \hline
20 &92.3 &496.2 &484.9&{8.23}\\ \hline
21 &85.9 &460.1 &457.5&{8.20}\\ \hline
25& 63.9 &298.1 &334.4&{8.32}\\ \hline
26& 45.1& 172.8& 232.0&{8.42}\\ \hline
27 &57.4 &234.1 &302.4&{8.37}\\ \hline\hline
    \end{tabular}
    \begin{tablenotes}
 \item $^*$: based on O32 calibration from \cite{Bian2018}.
 \item $^{(a)}$: [OII]$\lambda\lambda$3727,3729 outside the observed spectral range. Metallicity based on the code HII-CHI-mistry \citep{Perez-Montero2021}.
 \end{tablenotes}
\end{table*}

\subsection{Photometric catalogs}\label{sec:jwst-photometry}

We used version v0.51.2 of the CEERS Photometric Catalogs (Finkelstein et al. \textit{in prep}.). The catalog contains 101808 sources. The NIRCam images used are publicly available, and we refer the reader to  \cite{Bagley2022} for a complete description of the data reduction. We also make use of the HST data from CANDELS \citep{Grogin2011, Koekemoer2011}. For the CEERS pointings 1, 2, 3, and 6, the images are available in the Data Release 0.5\footnote{\url{https://ceers.github.io/dr05.html}}, while for the pointings 4, 5, 7, 8, 9, and 10, the images are available in the Data Release 0.6  \footnote{\url{https://ceers.github.io/dr06.html}}. 

A complete description of the photometric catalog will be presented in Finkelstein et al. \textit{in prep}. Briefly, the photometry was performed with SExtractor \citep[v2.25.0; ][]{Bertin1996} with F277W and F356W as the detection image. The fiducial fluxes were measured in small Kron apertures corrected by large-scale flux, following the methodology in \cite{Finkelstein2023}. 

Photometric redshifts were estimated with EAZY \citep{Brammer2008}, following the methodology in \cite{Finkelstein2023}. The code fits nonnegative linear combinations of user-supplied templates to derive probability distribution functions for the redshift based on the fit quality to the observed photometry for each detected source. The redshift is allowed to vary between $0-20$. For this catalog, new templates from \cite{Larson2022} were included to improve the goodness of the fit in high-$z$ sources. This new set of templates includes models with low metallicity {(5\% solar)}, young stellar populations {(10$^6$, 10$^{6.5}$, 10$^7$ yr)}, and high ionization parameters {(logU$=-2$)}. In this paper, we use the best-fitting photometric redshift defined as \textit{za} in EAZY. 

\begin{figure}[!t]
    \centering
\includegraphics[width=\columnwidth]{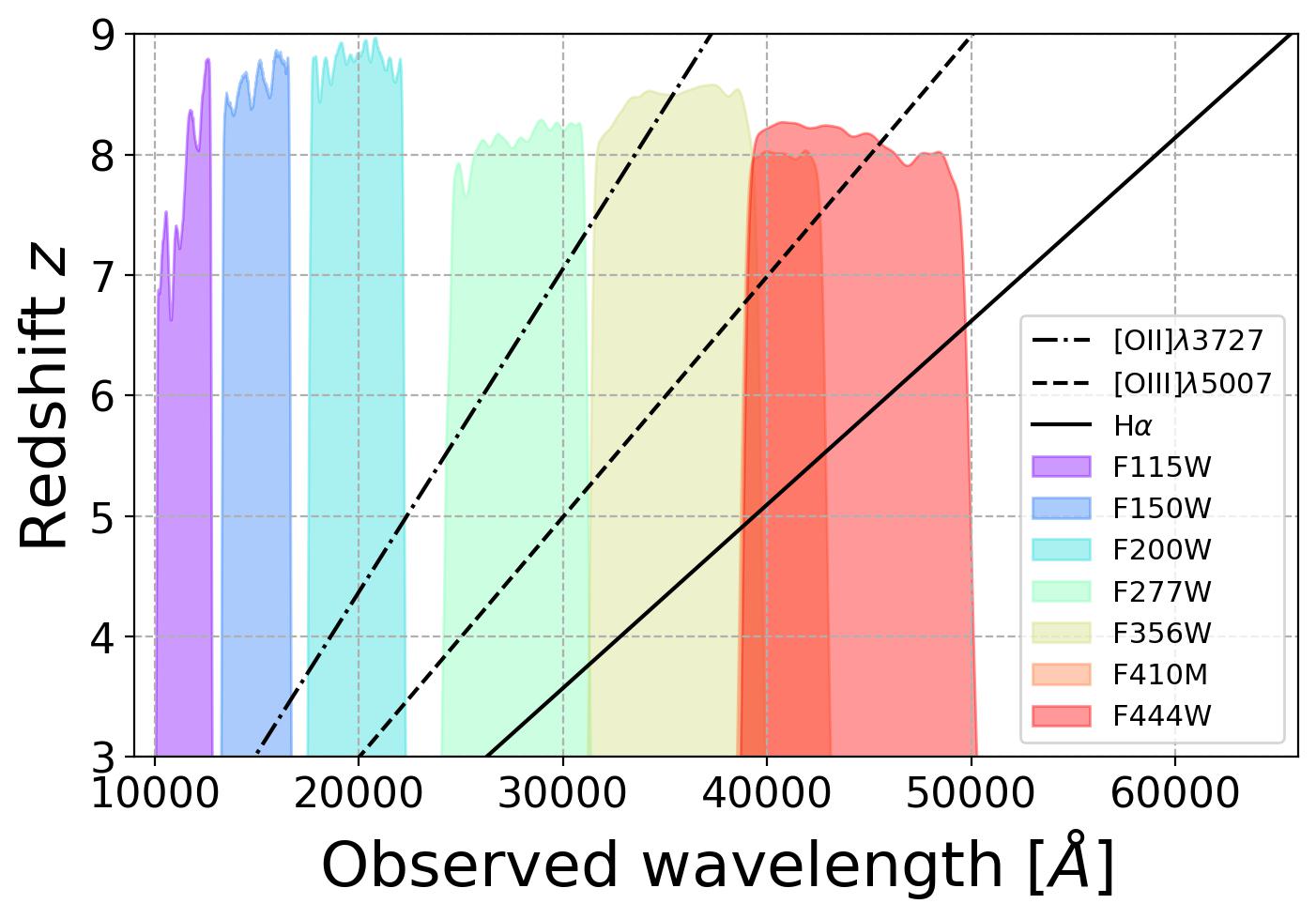}
    \caption{CEERS NIRCam filters. The black lines are the estimated observed wavelength at a given redshift for [OII]$\lambda$3727 (dotted-dashed), [OIII]$\lambda$5007 (dashed) and H$\alpha$ (solid). The filter response functions for NIRCam filters are arbitrarily scaled.}
\label{fig:wave-redshift}
\end{figure}

In this paper, we use CEERS NIRCam photometry and synthetic observations of templates of local EELGs to find such systems at $4\lesssim z<9$. In this redshift range, bright emission lines such as [OIII]$\lambda$5007 and H$\alpha$ may fall in the filters F277W, F356W, F410M, and F444W (see Fig. \ref{fig:wave-redshift}). We note that at $6.6 \lesssim z< 9$, H$\alpha$ redshifts beyond the bandpass of F444W, and then in this redshift range the selection is based only on [OIII]+H$\beta$ emission. The templates of local EELGs are described in the following Sec. \ref{sec:templates}.

\subsection{Spectroscopic data}\label{sec:spectro_CEERS}
The CEERS survey also includes six NIRSpec pointings, numbered p4, p5, p7, p8, p9, and p10. Each of these pointings has observations with the three NIRSpec medium resolution (G140M, G235M, and G395M) gratings and with the low-resolution Prism. Two more fields, p11 and p12, were observed with the prism in February 2023 because the prism observations p9 and p10 were severely impacted by a short circuit. The grating set covers from 0.97–5.10$\mu$m with a resolving power $R=\lambda / \Delta\lambda$ of $\sim$1000, while the prism covers from 0.60-5.30$\mu$m with a resolving power of $30 < R < 300$, depending on the wavelength. More objects can be observed simultaneously with the prism thanks to the shorter length (in pixels) of the prism spectra.

We adopt the NIRSpec data produced by the CEERS collaboration using the STScI JWST Calibration Pipeline\footnote{\url{https://github.com/spacetelescope/jwst}} \citep{Bushouse2022}.
Specifically, we use the JWST pipeline to perform the standard reductions, including the removal of dark current and bias, flat-fielding, background, photometry, wavelength, and slitloss correction for each exposure.
We also perform additional reductions to remove the $1/f$ noise and the snowballs.
The 2D spectra of each target are then rectified and combined to generate the final 2D spectra.
The details of the data reduction are presented in \citet[][]{ArrabalHaro2023} and Arrabal Haro et al. \textit{in prep.} 


\begin{figure*}[!h]
    \centering
    \includegraphics[width=\textwidth]{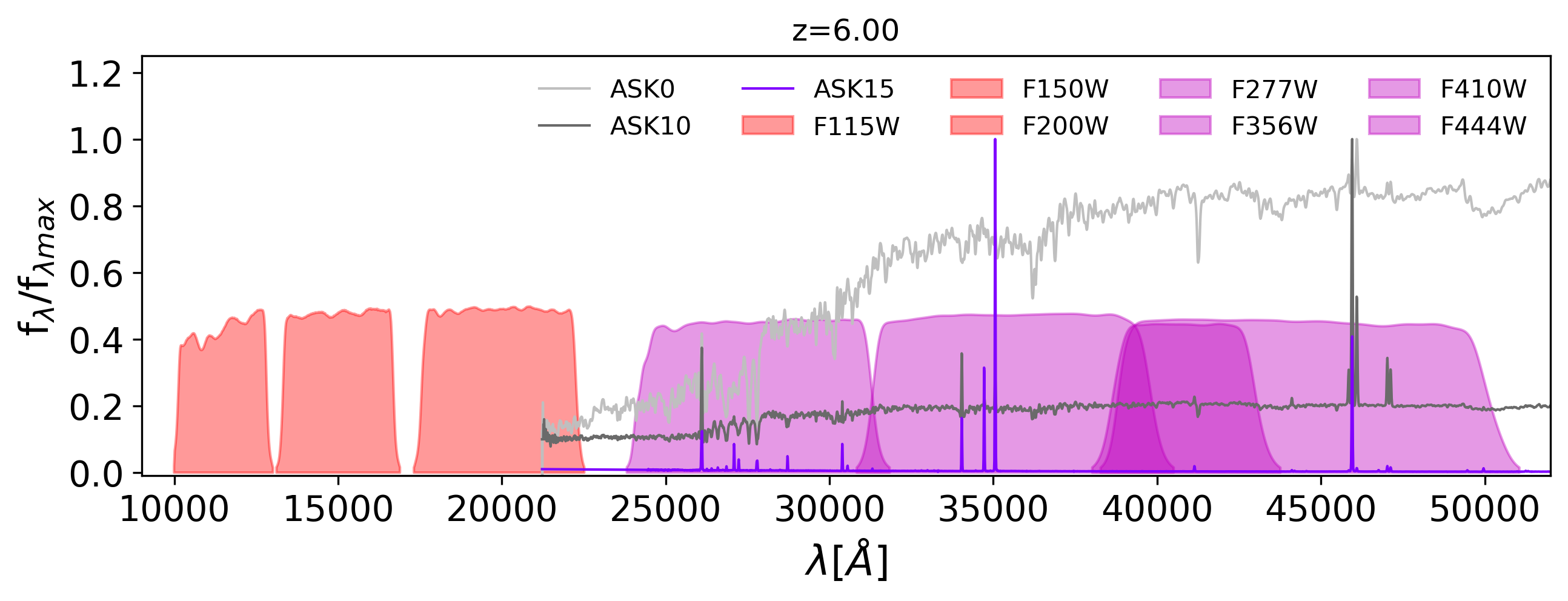}
    \caption{Illustration of the ASK templates at $z\sim 6$ in the NIRCam filters (\href{https://youtu.be/ltU8i_bsK4U}{Online movie}). We illustrate which filters are considered to estimate the NIRCam color due to the limited wavelength coverage of the templates. In red are the NIRCam filters where the templates are outside the filter, while in purple are the filters where we estimate the NIRCam colors. Three ASK templates are illustrated as examples (ASK0, ASK10, and ASK15). }
    \label{fig:templates_redshift}
\end{figure*}

\subsection{Empirical templates of EELGs}\label{sec:templates}

To create synthetic NIRCam observations of EELGs, we use the templates of the Automated Spectroscopic K-means-based (ASK) classes presented in \cite{SA2010}. According to this scheme, the $\sim$ one-million SDSS-DR7 galaxies with an apparent magnitude brighter than 17.8 can be classified in only 28 ASK classes based exclusively on the features and shape of their rest-frame and normalized by the $u$ magnitude of the optical spectrum.

The classification details are described in \cite{SA2010} but very briefly; to classify a galaxy, the spectrum is considered as a multidimensional vector, and it is assumed that the vectors are clustered around several cluster centers, known as classes. Then, the algorithm works iteratively to assign each rest-frame spectrum to the nearest class and then the class template as the average over all the class members. In the end, the algorithm finds the number of classes and their templates and assigns to each galaxy spectrum one of the classes according to a certain probability. All the galaxies in a class have very similar spectra; their average spectrum is the template spectrum of the class. We note that each template has a rest-frame wavelength coverage from $\sim$3000 to $\sim$9000\r{A}.

These ASK classes are labeled according to their $u - g$ color, from the reddest, ASK 0, to the bluest, ASK 27. Most ($\sim$99\%) galaxies in the SDSS-DR7 were classified into only 17 ASK major classes, with 11 additional minor classes including the remaining $\sim$1\%. \cite{SA2012} show that these rare classes correspond to metal-poor starbursts, and strong emission lines dominate their spectra. These minor classes, in particular, the classes ASK\,15, ASK\,17, ASK\,20, ASK\,21, ASK\,25, ASK\,26 and ASK\,27 show EW([OIII]$\lambda5007$) and EW(H$\alpha$)$>170$\r{A} in both lines (in Table \ref{tab:ask_classes}  we list our measurements using LiMe\footnote{\url{https://lime-stable.readthedocs.io/en/latest/}} which is a library that provides a set of tools to fit lines in astronomical spectra). The complete sample of local EELGs in the minor ASK classes is presented in \cite{Perez-Montero2021} and a more detailed analysis of the sample and templates will be presented in a future paper (Amorín et al. \textit{in prep}.). {We also include in Table \ref{tab:ask_classes} the gas-phase metallicity associated with each template derived using the O32 calibration from \cite{Bian2018}, which is the methodology that will be used in the following section \ref{sec:metallicity_gas}. We note that the templates are representative of metal-poor systems with metallicities ranging from 0.25-0.54Z$_{\odot}$}. We use the templates of these minor classes corresponding to EELG to estimate the expected colors of EELGs in the CEERS NIRCam photometry at $z>4$.

\subsection{Synthetic NIRCam observations}\label{sec:jwst:synt_nircam}
\label{sec:sintetico_nircam}

A simple and effective way of searching for EELGs using broad and medium band photometry is by using color selections, i.e., identifying the extreme colors caused by extreme emission lines producing excess up to $\sim 2$ mag in neighboring filters \citep[e.g.][]{Withers2023}.

Each spectrum of the 28 ASK classes was taken to a common wavelength range from 3030 to 9000\r{A}. We note that the ASK 15 has the shortest spectral range from 3747 to 8800\r{A}. In this case, we extrapolate the spectrum considering the continuum slope. Each spectrum was then redshifted to $z = 2 - 10$ (with $\Delta z=0.05$ step sizes). Synthetic NIRCam observations for the seven filters observed in CEERS were created for each redshift to search for broad and medium band color excesses driven by strong [OIII] + H$\beta$ and H$\alpha$ emission. The mean density flux of each template in each filter for each redshift was calculated by  
$f_{\nu}\propto\dfrac{\int \lambda T f_\lambda d\lambda}{\int \frac{T}{\lambda} d\lambda}$
where $T$ is the transmission curve of each filter, and f$_\lambda$ is the normalized template spectrum. In this way, the color in AB mag is given by $\rm{color}_{\rm 12}=-2.5[\log(f_{\nu 1})-\log(f_{\nu 2})]$. Given that the templates have a limited spectral range, our synthetic NIRCam observations are limited only to the filters where the templates fall completely in the filters. Due to this, we only estimate the magnitudes in the filters where the template is within 10\% of the filter transmission. An illustration of this is shown in Fig. \ref{fig:templates_redshift} for $z=6$, where the magnitudes for ASK templates are estimated only in magenta filters. A video for the complete $z$ range is available in this \href{https://youtu.be/ltU8i_bsK4U}{online movie}.

\begin{figure}[!h]
    \centering
    \includegraphics[width=\columnwidth]{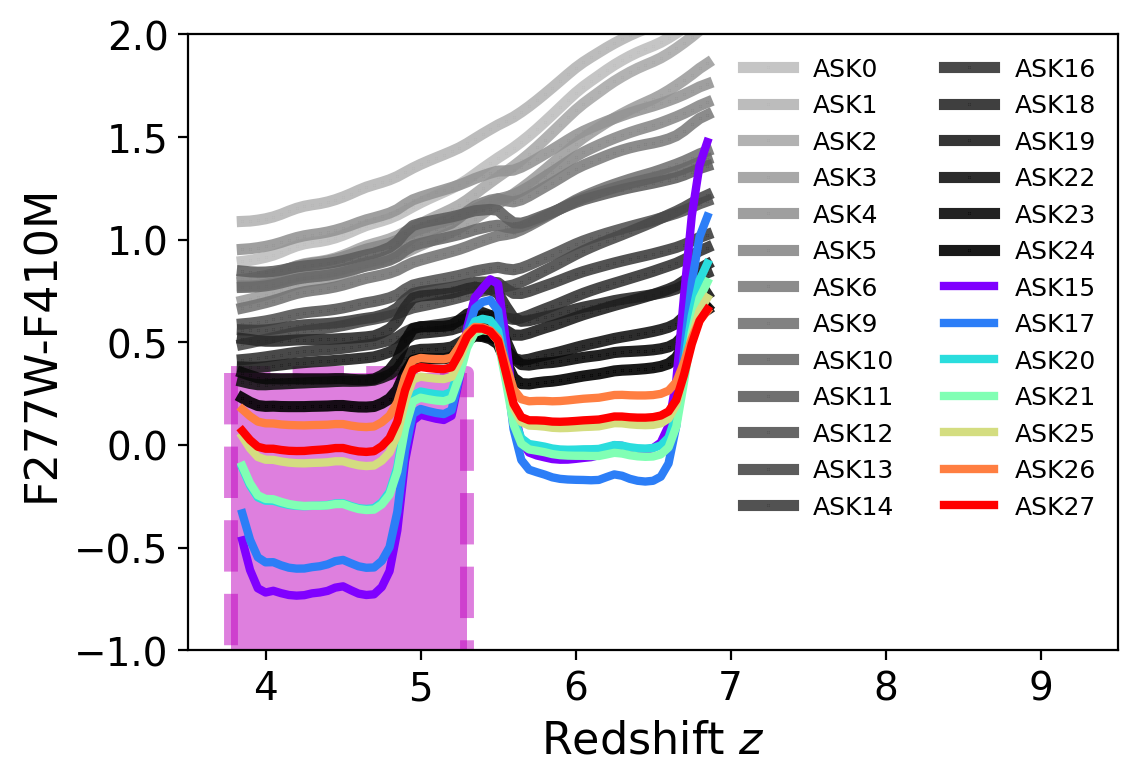}\\
    \includegraphics[width=\columnwidth]{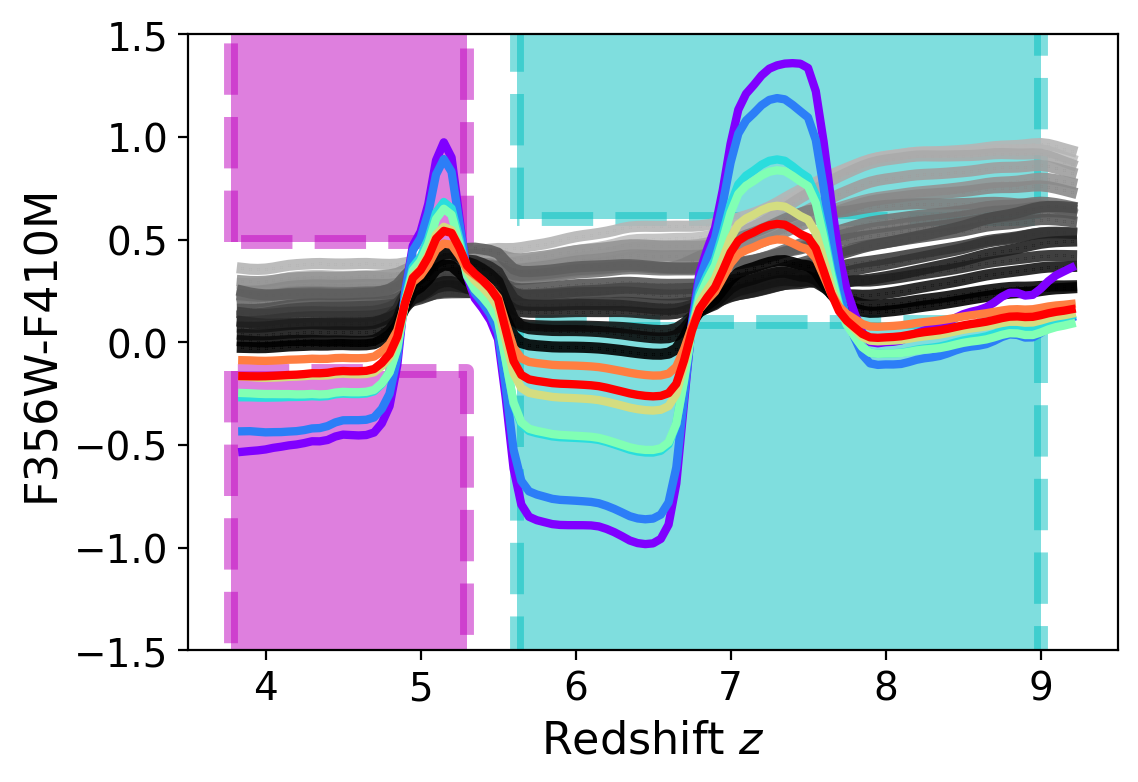}\\
    \includegraphics[width=\columnwidth]{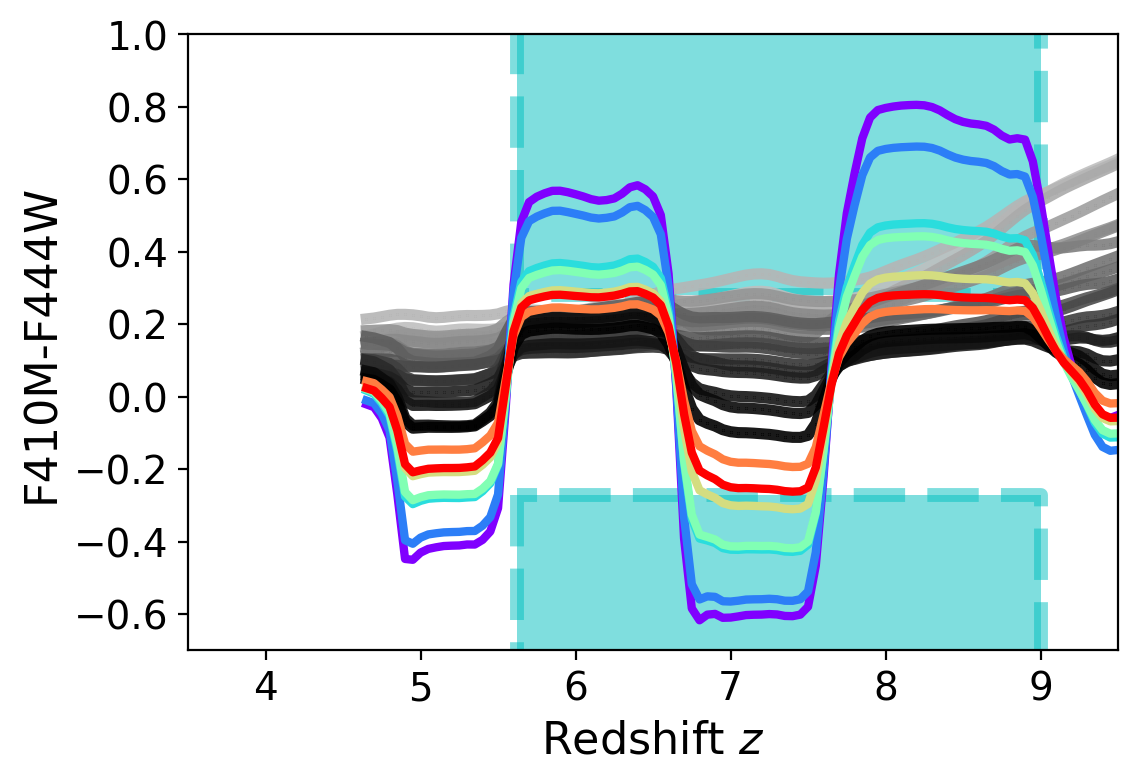}
    \caption{Redshift evolution of  NIRCam colors of ASK templates. The major ASK, i.e. non-EELGs,  classes are in gray-scale colors, according to the legend. The minor ASK classes, i.e. EELGs, are in colors according to the legend. The magenta and cyan-shaded regions are the $z$ ranges and color cuts used to select EELGs. }
    \label{fig:color_redshift}
\end{figure}

In Fig. \ref{fig:color_redshift}, we show the evolution with redshift of the colors used in this project, based on the synthetic NIRCam observations. With these results, we define color cuts based on the colors of the EELGs templates, which are separated from the other ASK classes in redshift windows. {We note that depending on the redshift of an EELG and the filters it is observed with, its colors can drastically change by up to $\pm$1 mag (Fig. \ref{fig:color_redshift}), and the effect is stronger with increasing EWs of emission lines. In that sense, an EELG may appear as an extremely blue or red source as strong emission lines move in and out of different filter transmission curves \citep[e.g.][]{Papaderos2012,Papaderos2023}.}

\subsection{Color-color diagrams}\label{sec:jwst_colordiagrams}
We define the color cuts based on the regions of redshift evolution of colors shown in Section \ref{sec:sintetico_nircam}. We use these color cuts to select candidates based on their position in color-color diagrams. While it is possible to identify EELGs by targeting emission from a single emission line
complex, our strategy requires strong emission in
both [OIII] + H$\beta$ and H$\alpha$. In this paper, we focus on the most extreme EELG candidates, and because of that, we consider only the templates ASK 15, 17, 20, and 21, which show EW([OIII]+H$\beta$)$>680$ \r{A}.

\begin{figure}[!h]
    \centering
    \includegraphics[width=\columnwidth]{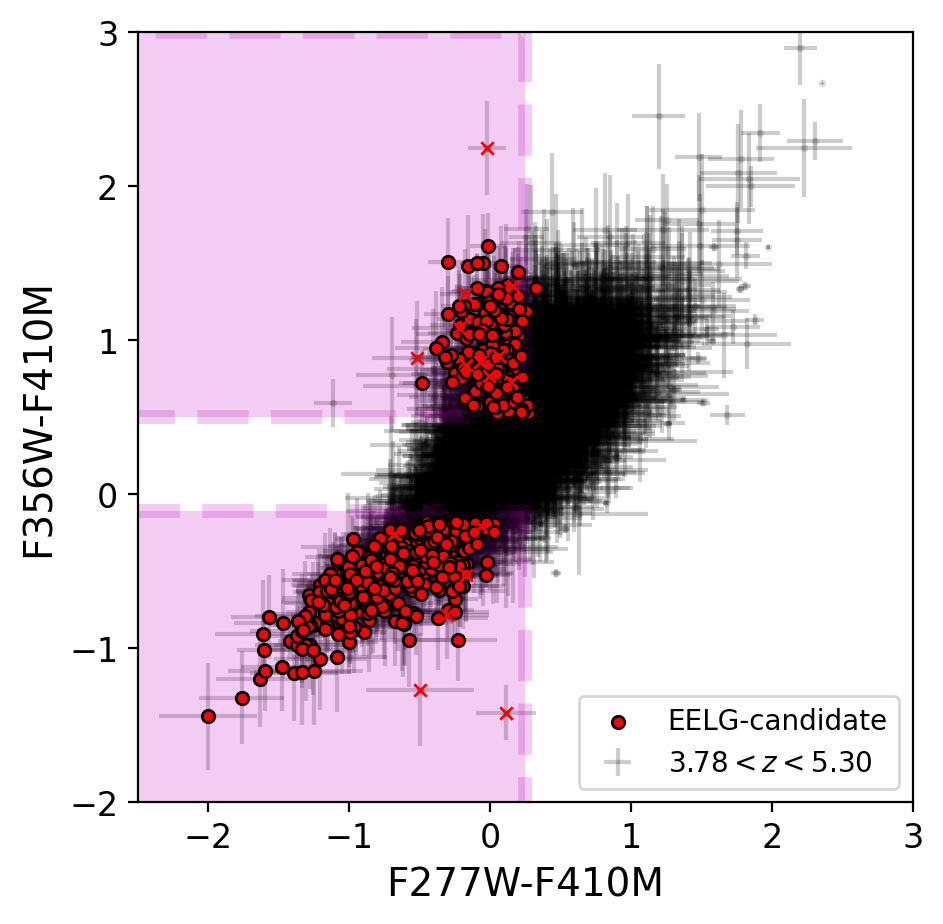}\\
    \includegraphics[width=\columnwidth]{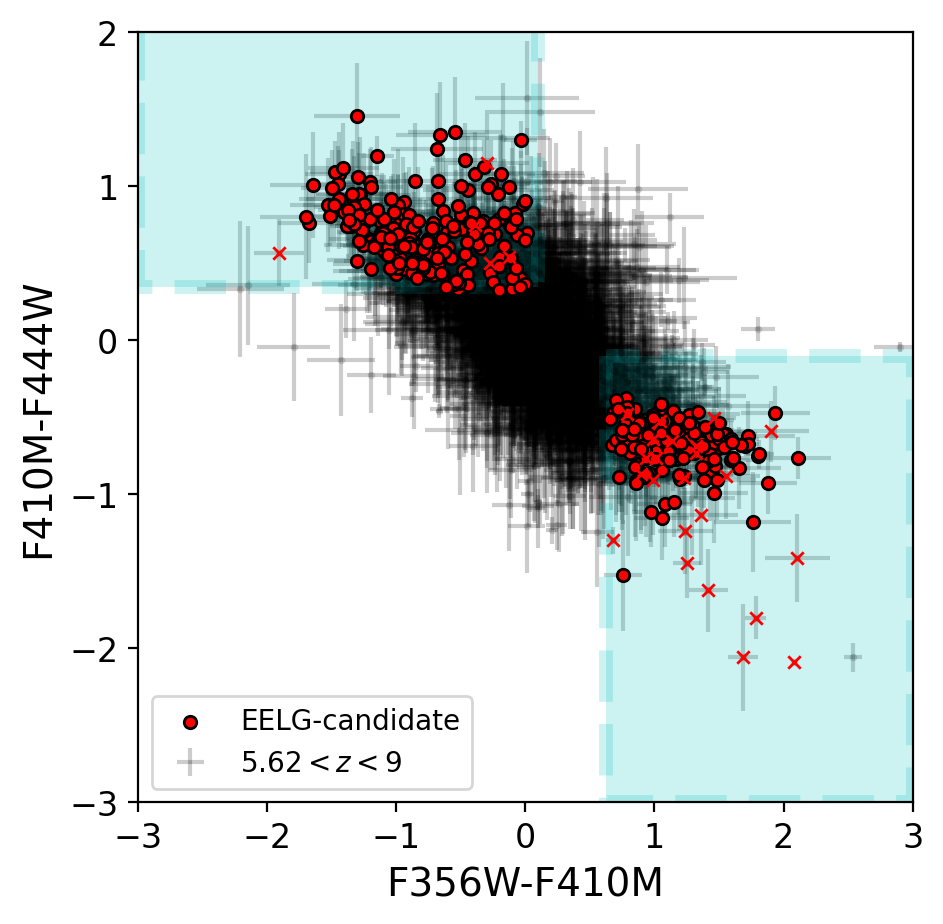}
    \caption{Color diagrams to select EELGs at each redshift range. The color cuts are marked in magenta (top panel) and cyan (bottom panel) shaded regions in each panel according to Eq. \ref{color_cut_z4} and \ref{color_cut_z5}, respectively. The selected EELG candidates are in red circles, while parent samples at similar redshifts are in black circles. The sources with red crosses were discarded due to contaminated photometry as explained in the main text.} 
    \label{fig:color-cut-diagrams}
\end{figure}

We focus on two redshift ranges (see color-color diagrams in Fig. \ref{fig:color-cut-diagrams}) which are selected due to the fact that EELGs can be distinguished from non-EELGs since bright emission lines enter or exit in a particular filter in the combination of colors. Additionally, we consider only sources with S/N$ >3$ in the photometric bands used in the color criteria. To remove objects with potential low-redshift solutions, we only considered sources whose $\int$P($z>3$)d$z>0.5$, ensuring that at least 50\% of the posterior probability is above $z=3$.  

{First, we consider sources whose photometric redshift is $z=3.78-5.3$ (top panel in Fig. \ref{fig:color-cut-diagrams}). In this redshift range, [OIII] falls in F277W, and H$\alpha$ falls in F356W or F410M. The numeric thresholds for the color cuts were chosen to include the colored more extreme EELG template curves in the color-$z$ space while excluding the grey curves for non-EELGs. We keep some regions where grey ordinary galaxies encroach upon the colored selection zones to allow a wider range in redshift since we are considering photometric redshifts. Based on our synthetic NIRCam color, we define the following color criteria:}
\begin{equation}
     \label{color_cut_z4}
      \left\{
	       \begin{array}{c}
F277W-F410M + \sigma(F277W-F410M)<0.35 \\
F356W-F410M + \sigma(F356W-F410M) <-0.14 \\
\lor  \\
F356W-F410M - \sigma(F356W-F410M)>0.49
	       \end{array}\right.
   \end{equation}

\begin{figure}[!t]
    \centering
\includegraphics[width=0.5\textwidth]{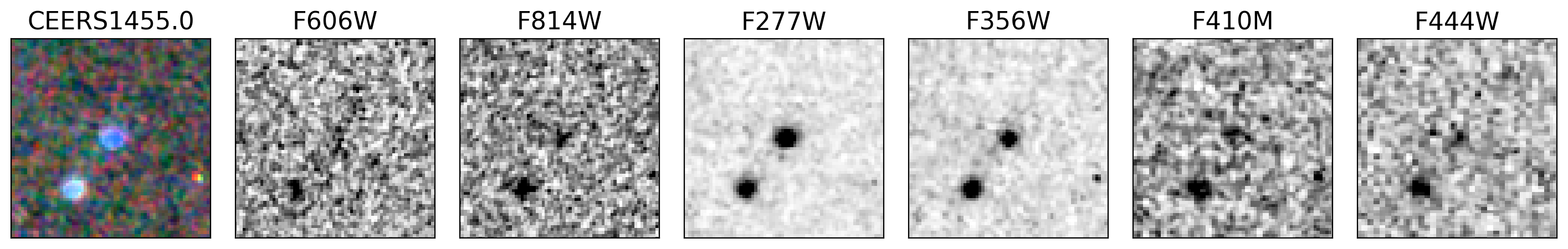}\\
\includegraphics[width=0.5\textwidth]{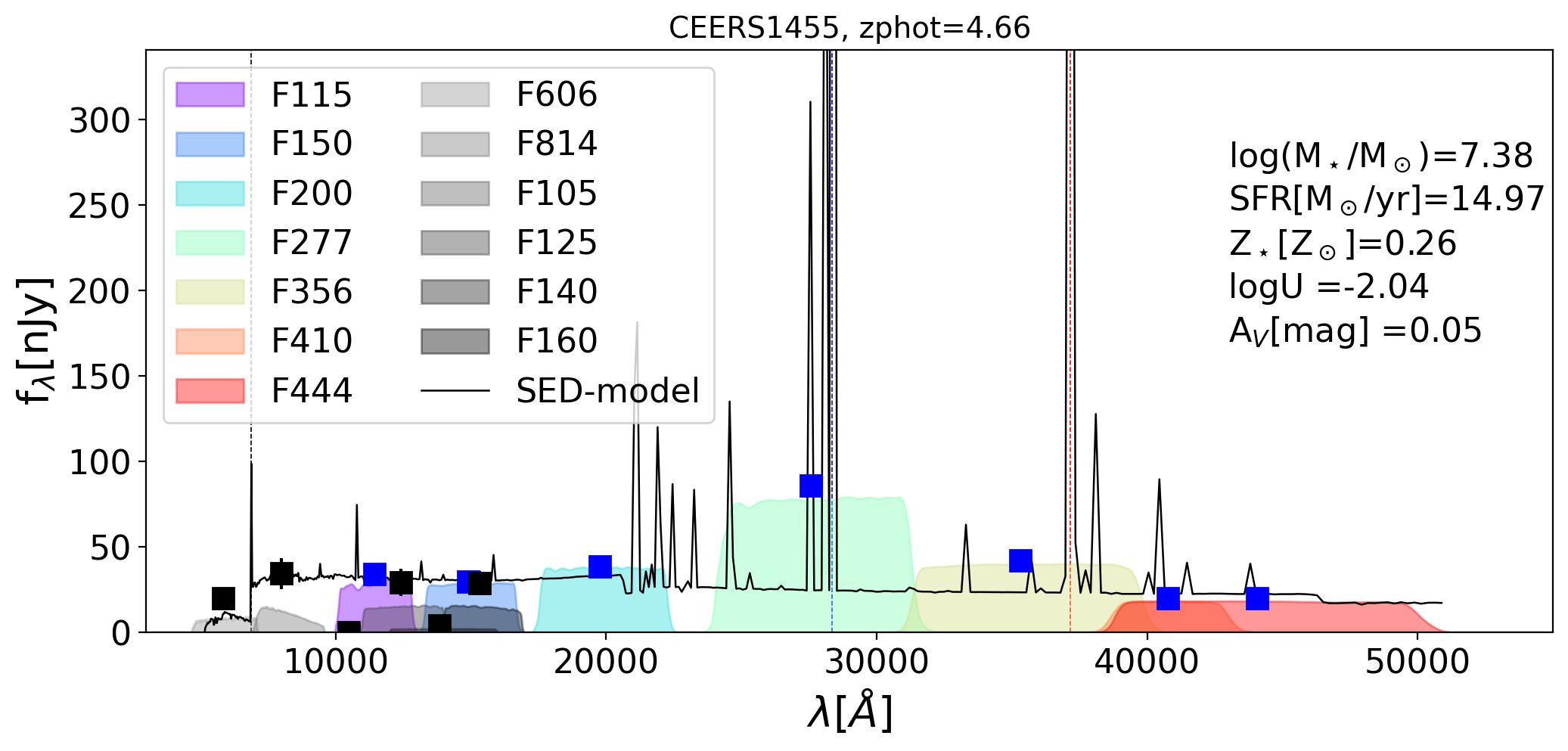}
    \caption{Images and SED of CEERS1455, an example of an EELG candidate at $z_{\rm phot}=4.66$. In the top panels, images (4''$\times$4'') of the galaxy in filters (from left to right) RGB image, HST/F606W, HST/F814W, JWST/F277W, JWST/F356W, JWST/F410M, and JWST/F444W. The RGB image is made with Red=F410M, Green=F356W, and Blue=F277W. On the bottom panel, SED of the galaxy. The blue (black) squares are the NIRCam (HST) photometric points. NIRCam (HST) filters are in rainbow-scale (gray-scale) colors, according to legend. The filter transmission is plotted in arbitrary units depending on the observed flux density. The black solid line is the SED model and the inset text summarizes its physical properties (see description in Sec. \ref{sec:SED-nircam}). The vertical dashed lines represent the position of Ly$\alpha$ (black), [OIII] (blue), and H$\alpha$ (red), according to their $z_{phot}$.}
    \label{fig:example_z4}
\end{figure}

In the color criteria, $\sigma$ refers to the observed color uncertainty. There are 4153 sources in this redshift range with S/N$>3$ in the three filters. 634 sources ($\sim 15$\%) satisfy the above conditions. An example of a candidate in this $z$ range is shown in Fig. \ref{fig:example_z4}.

{We then consider sources whose photometric redshift is in the range $z=5.62-9$ (bottom panel in Fig. \ref{fig:color-cut-diagrams}). In this range, [OIII] may fall in F356W, F410M, or F444W and H$\alpha$ in F410M or F444W}. We note that at $6.6 \lesssim z< 9$, H$\alpha$ redshifts beyond the bandpass of F444W, and then in this redshift range the selection is based only on [OIII]+H$\beta$ emission. Based on our synthetic NIRCam color, we define the following color criteria:\\
\begin{equation}
     \label{color_cut_z5}
      \left\{
	       \begin{array}{c}
F356W-F410M+\sigma(F356W-F410M)<0.1 \\

F410M-F444W-\sigma(F410M-F444W)>0.28 \\
\lor \\
F356W-F410M-\sigma(F356W-F410M)>0.6 \\

F410M-F444W+\sigma(F410M-F444W)<-0.27 \\
\end{array}\right.
   \end{equation}

\begin{figure}[!t]
    \centering
\includegraphics[width=0.5\textwidth]{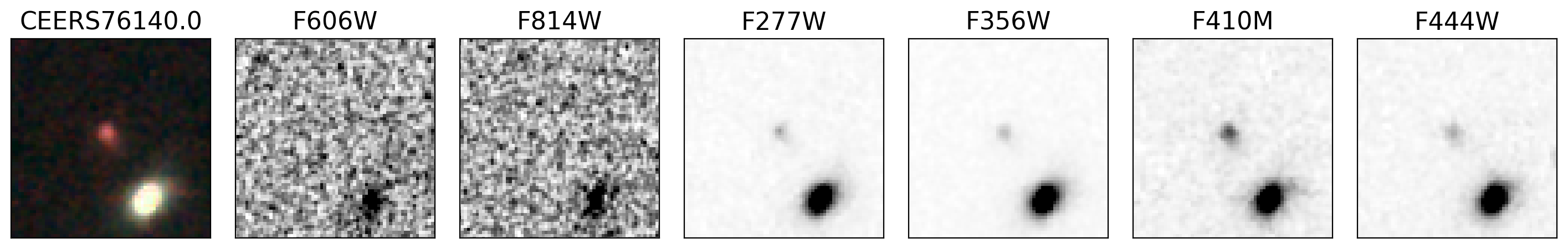}\\
\includegraphics[width=0.5\textwidth]{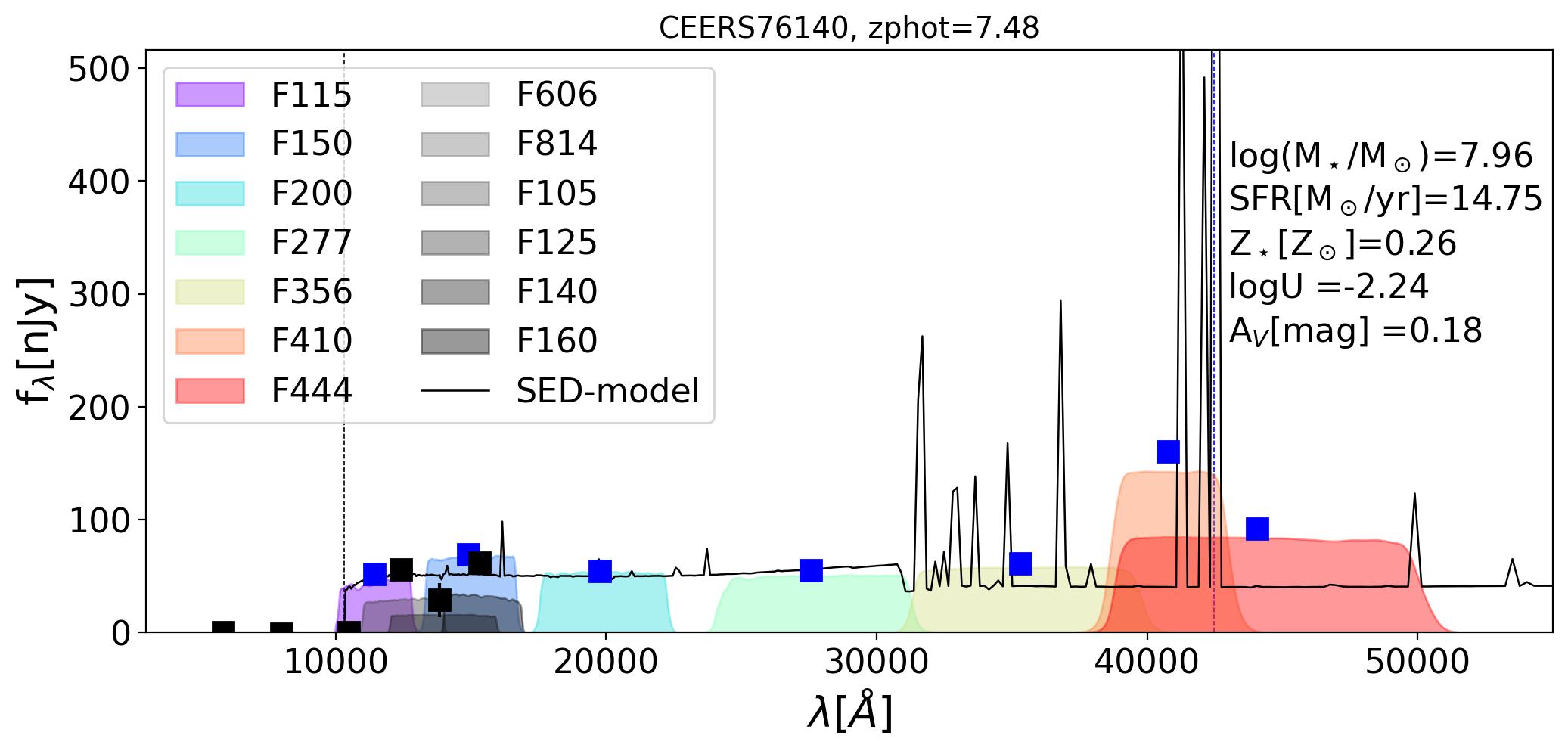}
    \caption{Images and SED of CEERS76140, an example of an EELG candidate at $z_{\rm phot}=7.48$. Symbols as in Fig.  \ref{fig:example_z4}.}
    \label{fig:example_z5}
\end{figure}

There are 2441 sources in this redshift range with S/N$>3$ in the three filters. 420 sources ($\sim 17$\%) satisfy the above conditions. An example of a candidate in this $z$ range is shown in Fig. \ref{fig:example_z5}. {We note that within this redshift range, for galaxies at $z>6.5$, the condition $\int$P($z>6.5$)d$z>0.5$ is satisfied which implies at least 50\% of the posterior probability is above $z=6.5$.}

With our method, we do not identify EELGs in the small interval between the two ranges, i.e., $5.3 < z < 5.62$, because [OIII] starts to enter the F356W filter and we can not distinguish both populations using this filter. We also note that we are considering the photometric redshifts in the selection of the candidates. Based on recent results \citep[e.g.][]{Davis2023}, the fraction of catastrophic redshifts ($\Delta z>3$) can be as high as 55\% along galaxies with signatures of extreme emission at similar redshifts. For this reason, we might be losing candidates in the selection if they have catastrophic redshifts. We restrict our selection using the photometric redshift to have the more secure candidates in the sample. {We also note that a complete validation of the photometric redshift is in preparation within the collaboration by comparing them with the entire sample with spectroscopic redshifts. For now, some comparisons have been performed in reduced samples \citep[e.g.][]{ArrabalHaro2023,Chworowsky2023} finding good agreement between photometric and spectroscopic redshifts with differences lower than 0.5. In Sec. \ref{sec:jwst_nirspec}, we will show a comparison with our sample of EELGs.}

\begin{figure}[!h]
    \centering
\includegraphics[width=0.5\textwidth]{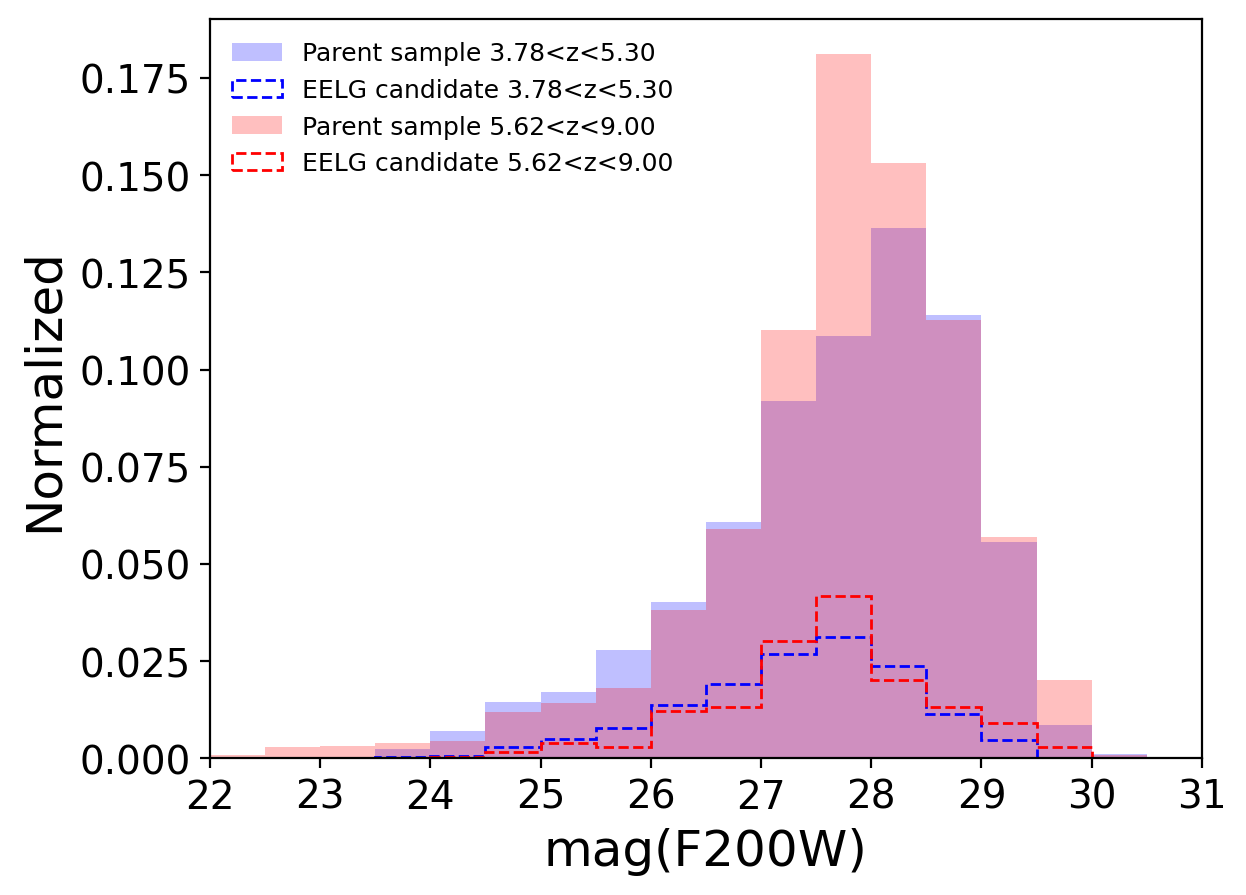}
    \caption{Distribution of the magnitude F200W for the sample of EELGs (unfilled histograms) and the parent samples at similar redshifts (filled histograms) in the two redshift ranges considered for the selection in blue and red, respectively. They are normalized to the total number of galaxies in each parent sample.}
    \label{fig:mag_200}
\end{figure}

\begin{figure}[!t]
    \centering
\includegraphics[width=0.5\textwidth]{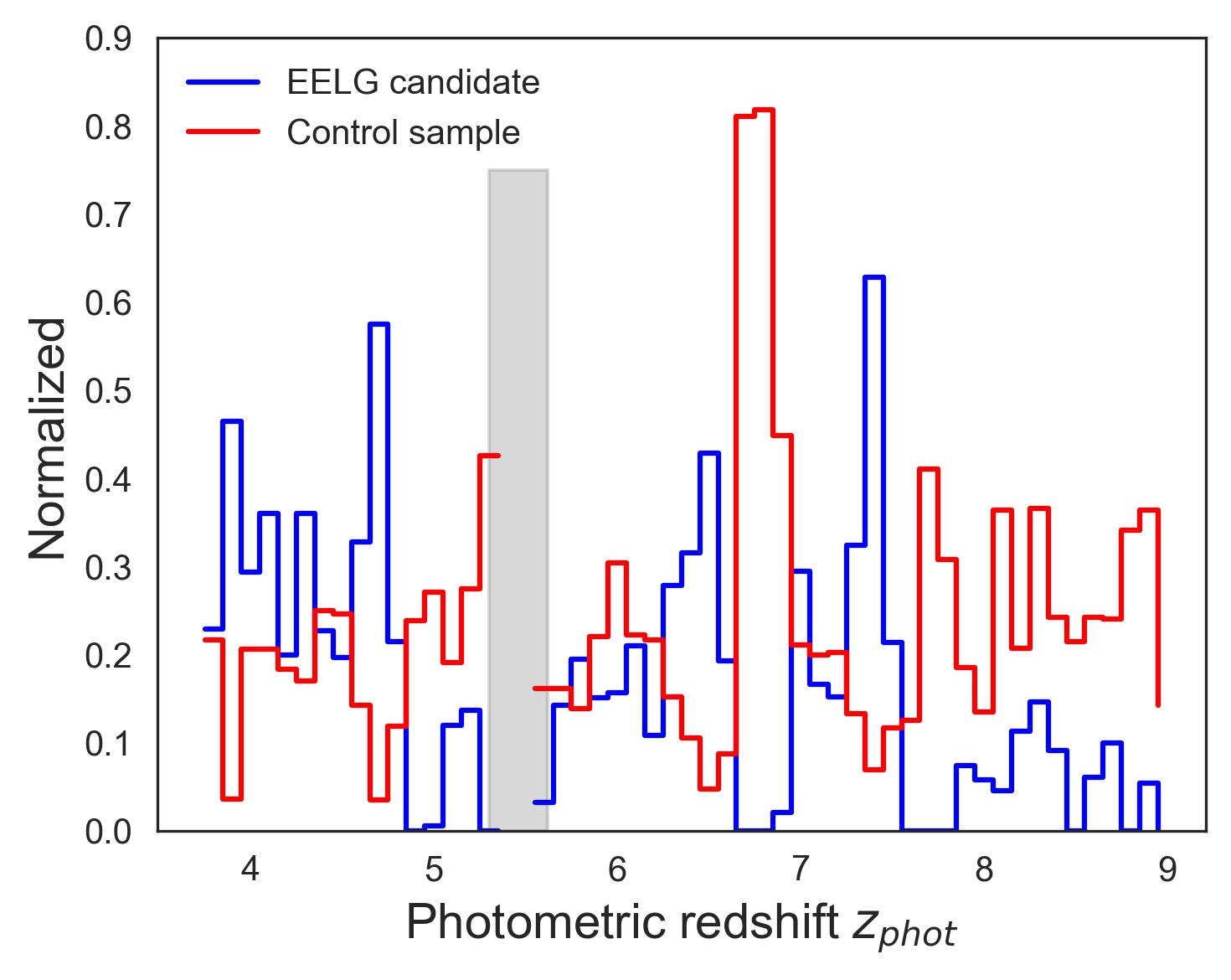}
    \caption{Distribution of photometric redshift of the final sample of EELG candidates (in blue) and the control sample (in red). The gray-shaded region is the gap between the redshift range for selection. The normalization is made for the total of galaxies in the parent sample at similar redshift and S/N.}
    \label{fig:EELG_redshift}
\end{figure}

\begin{figure*}[!t]
    \centering
\includegraphics[width=0.95\textwidth]{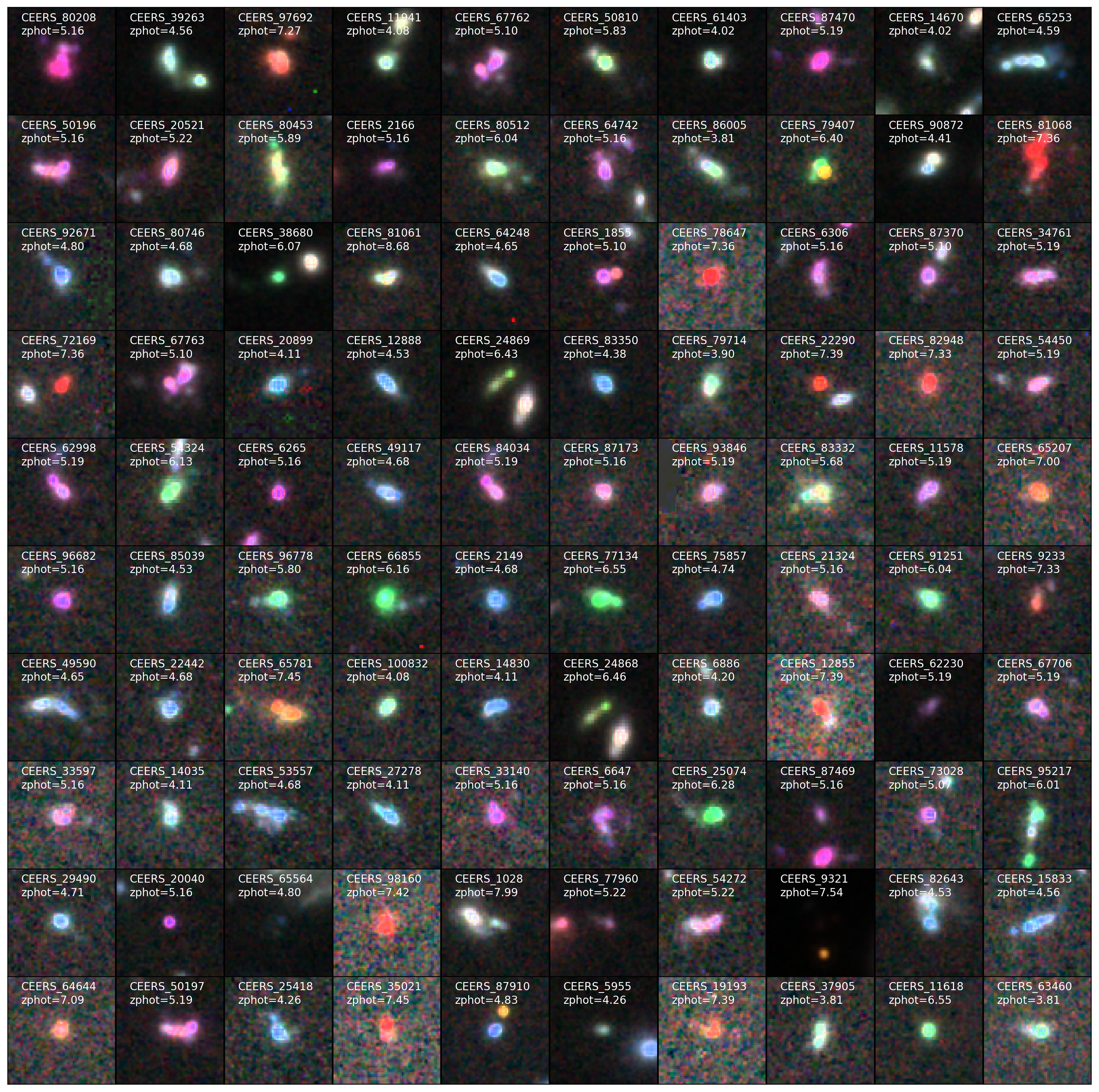}
    \caption{RGB (F410M-F356W-F277W) images of the 100th brightest F410M EELG candidates in our sample. Each image has a size of 4''x4''.}
\label{fig:images_stamps}
\end{figure*}
Our total sample of EELG candidates consists of 1054 galaxies. We highlight that by limiting our selection to galaxies with S/N$>3$ in their photometry, we are limiting our sample to galaxies with $<$29-30 mag in F200W and cover the magnitudes of their parent samples at similar redshifts and limited with the same S/N criteria in selecting filters (see Fig. \ref{fig:mag_200}). From these parent samples, we also define the control sample as the 1572 galaxies at the same redshift range and S/N that do not satisfy any selection color cuts in Eq. \ref{color_cut_z4} and \ref{color_cut_z5}. In Fig. \ref{fig:color-cut-diagrams}, the control sample is selected from black small circles- galaxies that do not reside in the hashed areas. To clean the control sample from galaxies with contaminated photometry, we additionally require that the Kron radius is larger than 1.6 pixels to avoid including hot pixels as sources. We selected this threshold after visual inspection of galaxies with Kron radius $<$1.6 pix which were mostly saturated pixels.  We aim to compare the physical properties of this control sample with the selected EELG candidates. 

We clean the sample of EELGs by performing a visual inspection of the images of the candidates to remove galaxies whose photometry might be contaminated. We do not perform a clean process for possible AGN in the sample, we only clean for sources with contaminated photometry. We remove 46 sources with saturated pixels and 8 sources in the spikes of nearby stars. Our clean sample contains 1000 EELG candidates \footnote{All the images of the EELG candidates are available in \url{https://github.com/mfllerena/EELGs/blob/main/images_EELG_CEERS.pdf}}. We flag (flag=1) a total of 60 candidates that are close to the edge of the detector, close to bright sources, and sources with low surface brightness. We keep these sources, which represent $\sim 6$\% of the sample of EELGs, in the analysis. The final redshift distributions of the EELG candidates and the control sample are shown in Fig. \ref{fig:EELG_redshift}. Each bin represents the percentage of the parent sample that is selected in the EELG and control sample. {We note that there is a peak of the number of galaxies in the control sample at redshift $z\sim 7$. In this redshift range, [OIII] starts to fall outside the F356W filter and inside the F410M and F444W filters. For this reason the colors tend to be zero and sources do not satisfy the selection criteria and most of the galaxies at this redshift range are in the control sample.}

We will also perform a further cleaning of this sample {of EELGs} based on the photometric EWs described in the following Sec. \ref{sec:photo-fluxes}. We validate our selection method by comparing it with available spectroscopy in the next Sec. \ref{sec:jwst_nirspec}.

The RGB (red=F410M, green=F356W, blue=F277W) images of the brightest F410M galaxies in our sample of EELGs are shown in Fig. \ref{fig:images_stamps}. The colors in RGB images are associated with redshift ranges where bright emission lines fall within filters. For example, blueish galaxies are at $z\sim 3.78-4.82$ where [OIII] falls in F277W and H$\alpha$ in F356W. In this redshift range, in particular galaxies at $z\sim 3.8$ may appear as greenish due to the low transmission of F277W compared with F356W for bright emission lines and/or H$\alpha$ being brighter than [OIII]. Purpleish galaxies are at $z\sim 4.82-5.31$ where [OIII] falls in F277W and H$\alpha$ in F410M. Green galaxies are at $z\sim5.62-6.63$ where [OIII] falls in F356W and H$\alpha$ in F444W. Redish galaxies are at $z\sim 6.63-7.68$ where [OIII] falls in F410M. For higher redshift galaxies ($z\sim 7.68-9$) [OIII] falls only in  F444W but [OII] falls in F356W and then they show light green colors.

We highlight that in the sample there are compact isolated sources (see for example CEERS29490, CEERS20040, CEERS64644 in Fig. \ref{fig:images_stamps}) but also nearby sources that are likely to be interacting (see for example CEERS53557, CEERS14830, CEERS54272 in Fig. \ref{fig:images_stamps}). Some galaxies in the sample look clumpy or chains of clumps (see for instance CEERS50196 or CEERS65253) dominated by one or more high surface brightness clumps, similar to EELGs at lower and intermediate redshifts \citep[e.g.][]{Amorin2015,Calabro2017}. There are cases with clear major companions that indicate they are currently experiencing a major merger (for instance CEERS65781 or CEERS62998 in Fig. \ref{fig:images_stamps}).

\section{Data analysis}

\subsection{Photometric fluxes}\label{sec:photo-fluxes}

\begin{figure}[!t]
    \centering
    \includegraphics[width=0.5\textwidth]{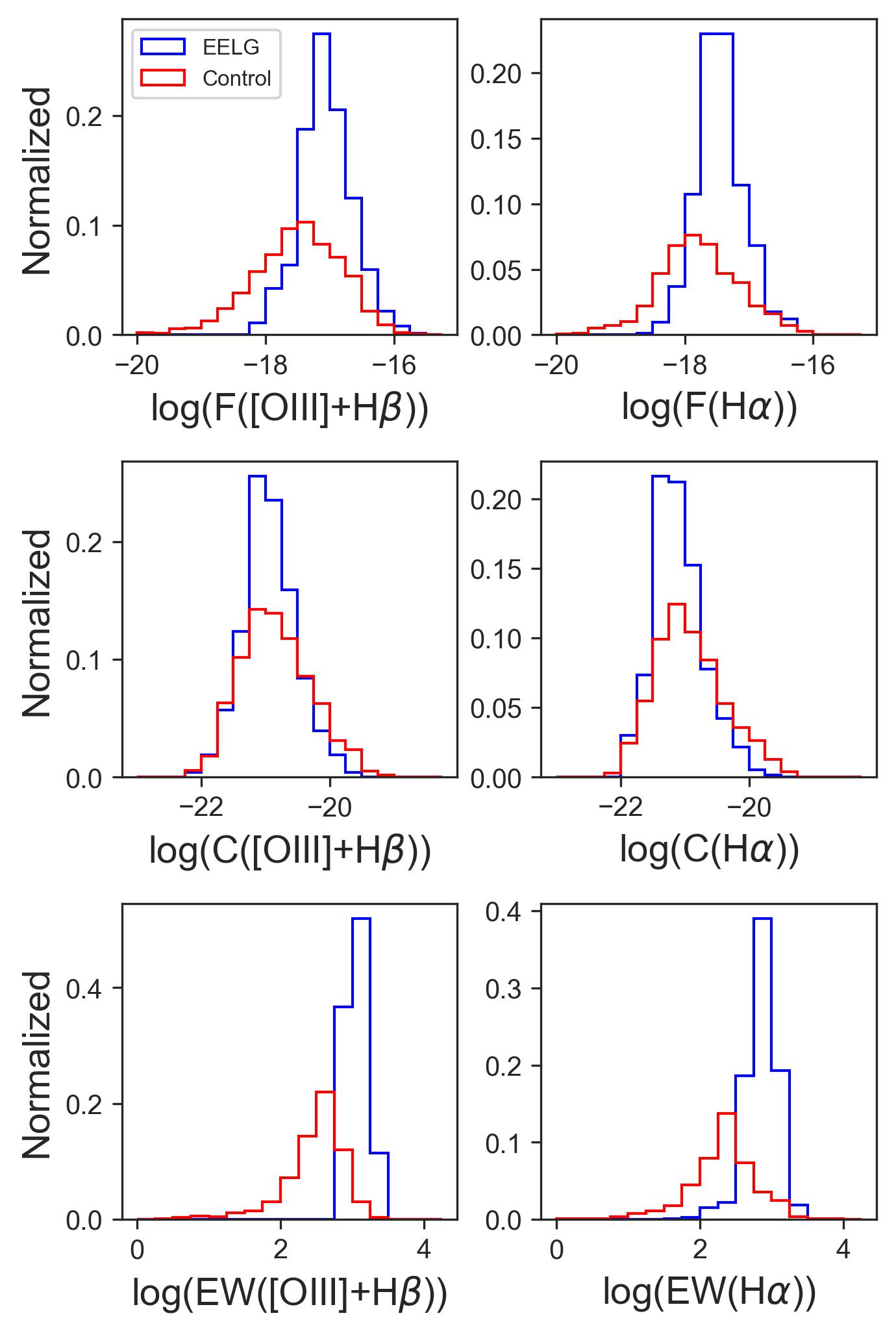}
    \caption{Photometric fluxes of [OIII]+H$\beta$ (left panels) and H$\alpha$ (right panels) for the sample of EELG candidates (in blue) and the control sample (in red). \textit{From top to bottom}: Integrated flux of the line in units of erg s$^{-1}$ cm$^{-2}$, continuum in the line wavelength in units of erg s$^{-1}$ cm$^{-2}$ \r{A}$^{-1}$, and rest-frame EW in units of \r{A}.}
    \label{fig:comparison_fluxes}
\end{figure}

We estimate the fluxes of [OIII]+H$\beta$ and H$\alpha$ based on the observed photometry, assuming the flux excess with respect to the continuum is due to the emission line that falls in the filter. We estimate the continuum by assuming that it follows a power-law   f$_{\lambda}\propto \lambda^{\beta}$ which matches the EELG templates. We consider all the NIRCam photometry observed in CEERS, excluding the bands where [OIII]+H$\beta$ and H$\alpha$ fall. We fit a linear model in log-log to estimate the $\beta$ slope using LMFIT \citep{NewvilleLMFIT}. We extrapolate the resulting model at the pivot wavelength of the filter to estimate the continuum of the line. The difference between the observed flux density and the estimated continuum is the flux of the line as F(line)=(f$_\lambda$-C$_\lambda$)$\Delta\lambda$, where $\Delta\lambda$ is the filter width and C$_\lambda$ is the continuum level. 

Based on the line fluxes, we estimate the rest-frame EW=$\dfrac{F(\rm{line})}{C_\lambda(1+z_{phot})}$. 
We note that these quantities might be overestimated since we are assuming all the flux excess is due to the bright emission line, but we are not considering other fainter lines that may contribute. For this reason, these quantities should be considered upper limits. We report our estimations in Table \ref{tab:photometry}.

We performed this estimation for the sample of EELG candidates and the control sample. We note that in our sample of EELGs we find 264 galaxies with photometric EW([OIII]+H$\beta$) $<680$\r{A}, which implies that our successful rate to detect EELGs with EW([OIII]+H$\beta$) $>680$\r{A} using the selection criteria in Eq. \ref{color_cut_z4} and \ref{color_cut_z5} is about 75\%. In the cases where EW([OIII]+H$\beta$) $<680$\r{A}, we note that the selection would improve if we added another condition in the selection, for example, the color F277W-F356W to probe the intense emission compared with the continuum for some redshift ranges. Or as already suggested in \cite{Davis2023}, the inclusion of an additional observational medium band filter such as F300M would benefit to sample the continuum. In the following sections, we only consider the 736 EELG candidates with photometric EW([OIII]+H$\beta$) $>680$\r{A} in the sample of EELGs.

The distributions can be found in Fig. \ref{fig:comparison_fluxes}. As expected, the sample of EELGs shows a higher fraction of galaxies with high fluxes of [OIII]+H$\beta$ and H$\alpha$ compared to the control sample. While the fraction of faint emitters ($\lesssim10^{-18}$ erg s$^{-1}$ cm$^{-2}$) is larger in the control sample. We also note that the continuum, which is the projection of the linear fit, is similar in both samples and no differences are found in the ranges. Regarding the EWs, the EELG sample shows a larger fraction of galaxies with larger values compared to the control sample. 

We highlight that the galaxies in the control sample with high EW([OIII]+H$\beta$) $>680$\r{A} show low EW(H$\alpha$) $<460$\r{A}, which is below the threshold of EWs for our selection color cuts. For this reason, these intense [OIII]+H$\beta$ emitters are not included in our sample of EELGs, since our selection criteria require that galaxies {at $z\lesssim 6.75$ (where H$\alpha$ falls still within the F444W filter)} are intense emitters of both [OIII]+H$\beta$ and H$\alpha$ above the thresholds defined by the color cuts. For this reason, there is an overlap between the sample of EELGs and the control sample regarding the high EWs. 

\begin{figure}[!t]
    \centering
\includegraphics[width=0.5\textwidth]{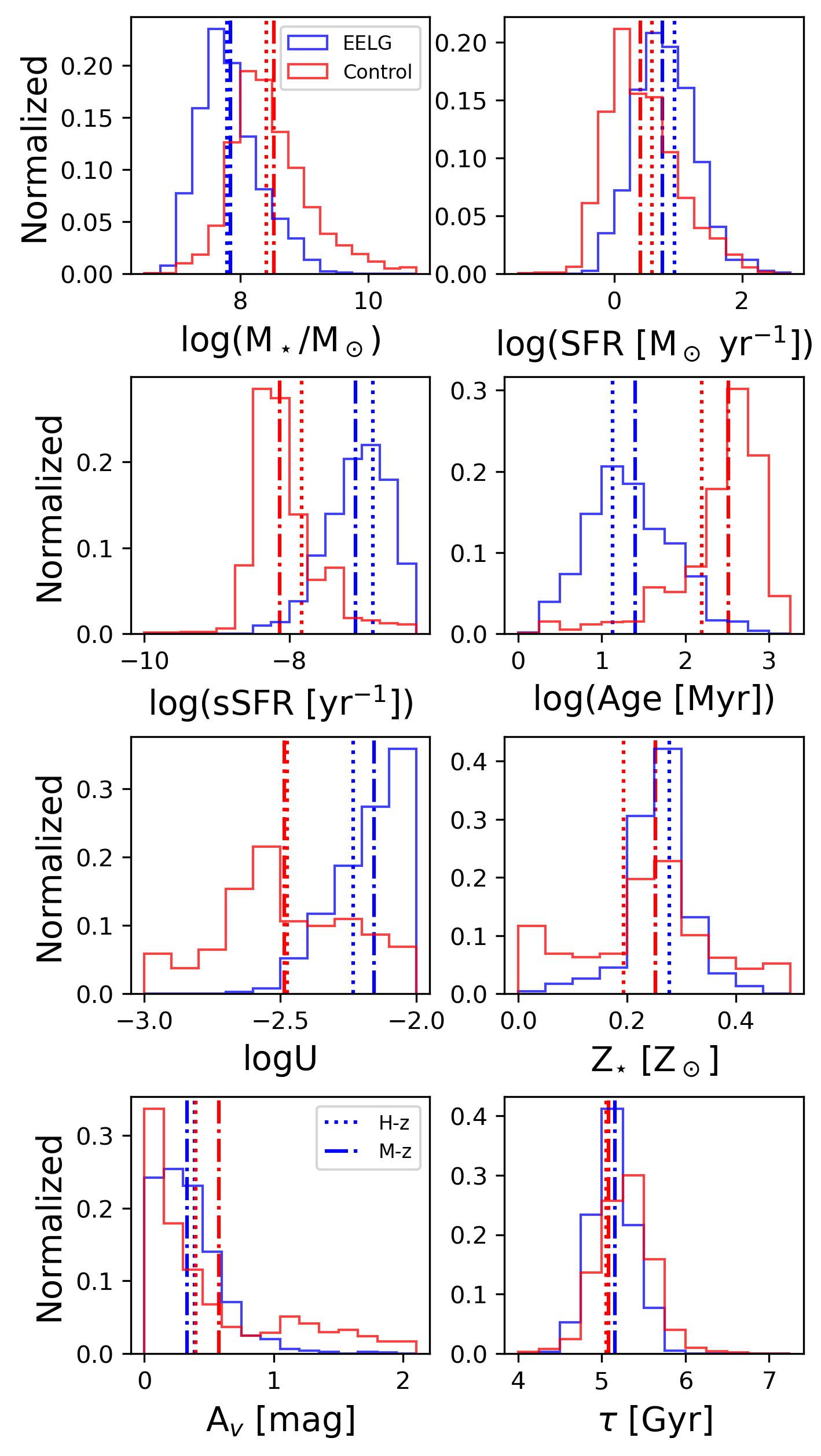}
    \caption{Distribution of the physical parameters based on the SED fitting for the sample of EELG candidates (in blue) and the control sample (in red). The vertical dotted (H-z sample) and dotted-dashed (M-z sample) lines are the mean values of each parameter.}
    \label{fig:SED_parameters_EELGs}
\end{figure}

\begin{table*}[!h]
    \caption{Photometry information for the first 10 EELGs candidates ordered by increasing CEERS ID number. We include their coordinates, photometric redshifts, and fluxes in the filters F277W, F356W, and F410M. Fluxes are in cgs units of erg cm$^{-2}$ s$^{-1}$. We also include the fluxes and rest-frame EWs of [OIII]+H$\beta$ and H$\alpha$ obtained from the photometry as described in Sec. \ref{sec:photo-fluxes}. In parenthesis, the uncertainties are reported. The complete table for the entire sample is available in the online version.}
    \label{tab:photometry}
\tiny{
\begin{tabular}{cccccccccccc}
   Id &     Ra &   Dec &  $z_{phot}$ &Flag$^*$&   F277W &   F356W &    F410M &        F([OIII]+H$\beta$) &       F(H$\alpha$) &     EW([OIII]+H$\beta$) &     EW(H$\alpha$) \\
   &     deg &   deg &   &&   nJy &   nJy &    nJy &        10$^{-18}$ cgs &       10$^{-18}$ cgs &     \r{A} &     \r{A} \\\hline
  14 & 214.924769 & 52.964053 &   4.78 &     1 &  70(4) &  40(5) &  25(6) & 11.9(1.2) &  2.6(1.1) & 1734(180) &  667(289) \\
 77 & 214.951387 & 52.982075 &   4.63 &     0 &  69(3) &  56(2) &  42(5) & 10.7(0.8) &  4.6(0.5) & 1466(111) &  956(107) \\
222 & 214.932836 & 52.968112 &   7.33 &     0 &   9(1) &   5(1) &  37(3) &  2.1(0.2) &        -- & 1011(120) &        -- \\
313 & 214.924964 & 52.962074 &   7.45 &     0 &   4(1) &   8(1) &  25(2) &  1.3(0.2) &        -- &  831(137) &        -- \\
400 & 214.955713 & 52.983425 &   6.49 &     0 &  14(1) &  35(1) &  15(2) &  4.5(0.3) &  3.2(0.5) & 1681(113) & 1630(249) \\
502 & 214.948019 & 52.977386 &   4.66 &     0 &  80(3) &  52(2) &  27(4) & 13.4(0.9) &  4.2(0.5) & 1724(116) &  944(106) \\
717 & 214.993544 & 53.008714 &   4.81 &     0 & 198(3) & 158(3) & 104(5) & 26.7(1.0) & 12.5(0.7) &  1073(41) &   888(46) \\
762 & 214.940365 & 52.970820 &   4.75 &     0 &  83(5) &  63(4) &  36(8) & 12.3(1.6) &  4.2(0.9) & 1316(171) &  713(156) \\
773 & 214.923827 & 52.959152 &   7.36 &     0 &     -- &  12(3) &  43(6) &  2.5(0.5) &        -- & 1044(220) &        -- \\
795 & 214.948392 & 52.975950 &   4.66 &     0 &  60(4) &  38(3) &  25(5) & 10.3(1.0) &  2.9(0.7) & 1843(188) &  846(197) \\\hline
\end{tabular}
\begin{tablenotes}
 \item $^*$: Quality flag of the photometry where 0 is reliable photometry and 1 is a source near a bright source, close to the edge of the detector, or with low surface brightness.
 \end{tablenotes}}
\end{table*}

\subsection{Physical parameters}\label{sec:SED-nircam}
To obtain the physical properties of the sample of EELGs and the control sample, we perform SED fitting with only JWST photometry to have homogeneous data. We consider the entire set of filters used in the CEERS surveys, which are F115W, F150W, F200W, F277W, F356W, F410M, and F444W. 
We used BAGPIPES \citep{Carnall2018} to estimate the physical parameters with the \cite{Bruzual_2003} stellar population models. We fixed the redshift to the photometric redshift. We consider a delayed exponential $\tau$-model for the SFH, where $\tau$ is the timescale of the decrease of the SFH. In the model, we consider the age ranging from 1Myr to the age of the Universe at the observed redshift. We allow the $\tau$ parameter to vary between 0.1 to 10 Gyr. We allow the metallicity to vary up to 0.5Z$_\odot$ freely. For the dust component, we consider the \cite{Calzetti2000} attenuation curve and let the A$_V$ parameter vary between $0-2$ mag. We also include a nebular component in the model, and we let the ionization parameter freely vary between $-3$ and $-2$. Some examples of the SED model are shown in Fig. \ref{fig:example_z4} and \ref{fig:example_z5}.

The distribution of the main obtained SED parameters is shown in Fig. \ref{fig:SED_parameters_EELGs}, which includes stellar mass, SFR, sSFR, age, logU, stellar metallicity, absolute attenuation A$_V$, and $\tau$.

We report the obtained parameters in Table \ref{tab:sed}. We find that our sample of EELGs shows stellar masses between 10$^{6.83}$\Msun~and 10$^{9.65}$\Msun~with a mean value of 10$^{7.84}$\Msun. They are actively forming stars with SFRs between 0.41 to 900 \Msun yr$^{-1}$, with a mean value of 13.64 \Msun yr$^{-1}$. This implies they show very high sSFR above 10$^{-8.47}$yr$^{-1}$ and up to 10$^{-6.1}$yr$^{-1}$, with a mean value of 10$^{-7.03}$yr$^{-1}$. We also find that these galaxies are young, with times after the onset of star formation of roughly 1.7-792 Myr, with a mean value of 45 Myr. Regarding the ages, we also tested an exponential $\tau$-model as SFH and we found a difference of 0.28 dex in the ages between both models which is a factor $\sim\times2$ larger ages with the delayed exponential model. 

Regarding the ISM properties of the EELGs, we find they show high ionization parameters with a mean value of logU$=-2.17$, with a saturation at logU$=-2$ which indicates harder ionizing spectra than those considered in the models are needed to model these sources. They show subsolar metallicities ranging from 0.008 to 0.44 Z$_{\odot}$ and a mean value of $ \sim $0.25Z$_{\odot}$. They also show low dust extinction, with A$_V$ values ranging from $\sim 0$ to 1.87 mag, but with a mean value as low as 0.34 mag. We note that we are considering a delayed $\tau$-model for the SFH, but the results of the SED fitting indicate $\tau$ values of $\sim$5.1 Gyr on average, which indicates that they are consistent with a constant SFH. 

As shown in Fig. \ref{fig:SED_parameters_EELGs}, where we compare the sample of EELGs with the control sample, we note that both samples show similar ranges of stellar metallicity and absolute attenuation with similar mean values. Regarding other parameters, we find that the EELG candidates show lower stellar masses and higher SFRs and sSFRs compared to the control sample. Comparing the mean values, we find that the sSFRs of the EELGs are on average $\sim$0.8 dex higher than the mean value of the control sample.
EELGs also show younger ages compared with the control sample which shows a mean value of 405Myr. The control sample also shows a lower value of logU with a mean value of $-2.48$. To statistically verify these differences we performed a Kolmog\'orov-Smirnov (K-S) test and we found that for all parameters in Fig. \ref{fig:SED_parameters_EELGs} the two samples are significantly different, i.e. they show a p-value $<0.05$. In the parameters where these differences are more significant, i.e. lowest p-values ($\sim 10^{-211}-10^{-113}$), are the sSFR, ages, and ionization parameter. While in the absolute attenuation, $\tau$-parameter, and stellar metallicity these differences are less significant with higher p-values ($\sim 10^{-20}-10^{-17}$) compared with the rest of the parameters. 

{We also split our samples of EELGs and control in two redshift bins: M-z for galaxies at $z\leq 6.5$ and H-z for galaxies at $z>6.5$ (see the vertical lines in Fig. \ref{fig:SED_parameters_EELGs} for their mean values) to check if there are differences between both populations. Regarding stellar mass, we find that both populations of EELGs show similar mean values. Regarding the SFR, we find that the M-z EELG galaxies show slightly lower ($\sim$ 0.18 dex) SFRs than H-z EELGs. This implies that the M-z EELGs show slightly lower ($\sim$ 0.2 dex) sSFRs than H-z EELGs. Similar trends are also observed in the control sample. Regarding the ages, the H-z EELGs show a mean value of 25Myr while the M-z EELGs show a value of 51Myr. Regarding the control sample, the M-z sample shows a mean age of 486Myr while the H-z sample shows 235Myr, which is expected due to the age of the universe. Even though we find differences in the mean values of these parameters when split by redshift, we note that the differences with the control sample are more significant than the differences with redshift.}

In summary, EELGs are younger, more star-forming, and have higher ionization parameters compared to the control sample. 

\begin{table*}[!h]
    \centering
    \caption{SED parameters for the first 10 EELGs candidates ordered by increasing CEERS ID number. We include their stellar mass, SFR, age, ionization parameter, stellar metallicity, Av. We also report the physical size in the F200W filter and $\Sigma_{\rm SFR}$ as described in Sec. \ref{sec:sizes}. In parenthesis, the uncertainties are reported. The complete table for the entire sample is available in the online version.}
    \label{tab:sed}
\tiny{
\begin{tabular}{c|c|c|c|c|c|c|c|c}
   Id &        log(M$_{\star}$/\Msun) &          SFR &      Age &      logU &          Z$_{\star}$  &A$_V$&r$_{\rm eff}$ (F200W)&$\Sigma_{\rm SFR}$\\
   &         &          \Msun yr$^{-1}$ &      Myr &       &          Z$_{\odot}$ &mag& kpc&\Msun yr$^{-1}$ kpc$^{-2}$\\\hline
 14 & 7.47(0.14) &  4.5(1.7) &   12(10) & -2.1(0.1) & 0.24(0.09) & 0.16(0.10) & 0.23(0.03) &  13.39(6.50) \\
 77 & 7.91(0.09) &  6.9(1.7) &   23(14) & -2.1(0.1) & 0.25(0.10) & 0.77(0.12) & 0.57(0.07) &   3.38(1.17) \\
222 & 7.15(0.10) &  8.2(2.7) &     2(1) & -2.2(0.2) & 0.32(0.11) & 0.37(0.19) &         -- &           -- \\
313 & 7.84(0.53) &  5.7(3.4) &   28(95) & -2.3(0.2) & 0.30(0.13) & 1.30(0.43) &         -- &           -- \\
400 & 8.02(0.17) & 13.4(4.3) &   16(13) & -2.2(0.1) & 0.25(0.15) & 1.39(0.19) &         -- &           -- \\
502 & 7.52(0.08) &  6.1(1.5) &    10(4) & -2.1(0.1) & 0.25(0.08) & 0.24(0.08) &         -- &           -- \\
717 & 8.05(0.07) & 12.2(1.0) &    19(4) & -2.2(0.1) & 0.15(0.01) & 0.11(0.03) &         -- &           -- \\
762 & 8.28(0.27) &  3.6(2.5) & 125(376) & -2.2(0.1) & 0.26(0.11) & 0.41(0.19) &         -- &           -- \\
773 & 7.84(0.55) &  4.2(4.5) &  35(149) & -2.4(0.2) & 0.29(0.13) & 0.61(0.56) &         -- &           -- \\
795 & 7.54(0.21) &  4.3(1.8) &   16(20) & -2.1(0.1) & 0.25(0.09) & 0.39(0.12) & 0.48(0.26) &   2.94(3.45) \\\hline
\end{tabular}}
\end{table*}

\subsection{Physical sizes and SFR surface density}\label{sec:sizes}

To measure the sizes of the EELG candidates and the control sample, we use the Galfit catalogs (v0.52) of the collaboration (McGrath et al. \textit{in prep}.). Galfit \citep{Peng2002,Peng2010} was run for sources with F356W$<$28.5 mag using background-subtracted mosaics. Most of the EELG candidates are in the catalog (84\% of the sample) independent of the quality of the fit. The photometric catalogs were used for making first guesses on source location, magnitude, size, position angle, and axis ratio. The Kron radius was used to determine an appropriate image thumbnail region for Galfit to fit. All galaxies within 3 magnitudes of the primary source, but no fainter than 27~mag 
that fell within the thumbnail region were fit simultaneously. Galaxies that were not fit simultaneously were masked during the fitting process using the 
segmentation map. This procedure was performed for both F200W and F356W images. Here we only considered the F200W results since they are tracing mostly the rest-frame$<$1750-4680\r{A} (depending on redshift) and therefore they represent well the size of the young stellar populations. In this way, we minimize the contribution from the ionized gas with bright emission lines. As a caveat, we note that the nebular continuum might be significant in EELGs and could contribute to the sizes that we use in this paper. We also note that the filter F200W may have a contribution of emission lines such as [OII]$\lambda3727$ for the galaxies at the lowest redshifts considered ($3.8<z<5$) which might lead to overestimations of their rest-UV sizes.  

\begin{figure}[!t]
    \centering
    \includegraphics[width=0.5\textwidth]{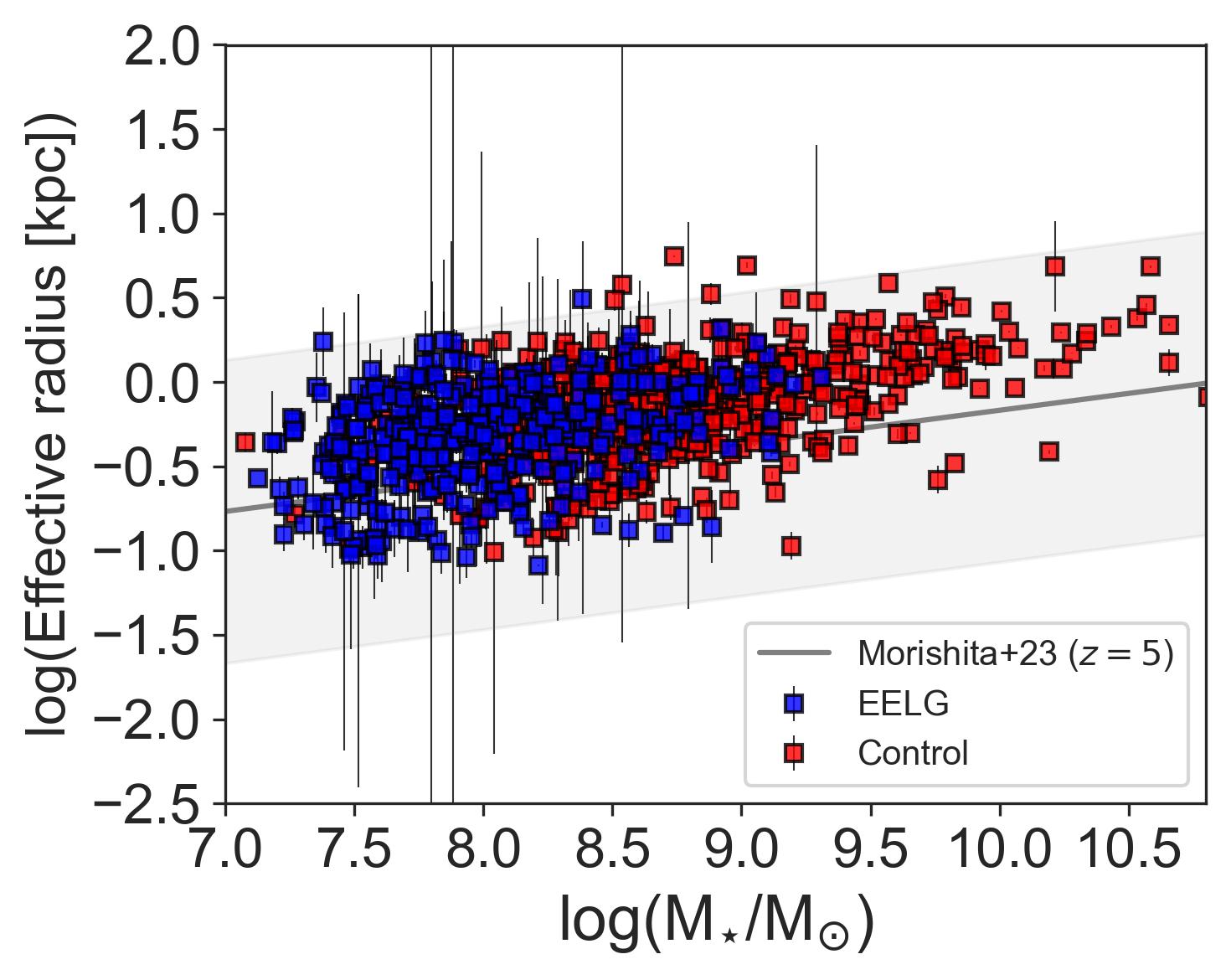}
    \caption{Size-mass relation for the EELG (in blue) and the control (in red) samples. The grey solid line is the size-mass relation at $z=5$ from \cite{Morishita2023} and the shaded region the 3$\sigma$-scatter.}
    \label{fig:size-mass}
\end{figure}

We only consider 360 sources with good fit quality (so-called Flag=0). In Fig. \ref{fig:size-mass}, we show that the galaxies lie within 3$\sigma$ of the size-mass relation at $z\sim5$ reported in \cite{Morishita2023} for rest-UV sizes. The subsample of EELGs shows effective radii ranging from 80pc to 3kpc, with a mean value of 0.57kpc. Regarding the control sample, we find that 74\% of them are in the Galfit catalog, but only 526 of them have good fit quality. This subsample is also displayed in Fig. \ref{fig:size-mass}. They show a slightly higher mean effective radii of 0.94kpc.

Based on the F200W sizes and the SFRs from the SED model, we estimate the projected SFR surface density as $\Sigma_{\rm SFR}=\dfrac{\rm SFR_{SED}}{2\pi r_{\rm eff}^2}$, where $r_{\rm eff}$ is the effective radius as the size in the F200W image. We find that the galaxies in our sample of EELGs show a wide range of $\Sigma_{\rm SFR}$ values ranging from 
0.14 to 930 \Msun yr$^{-1}$ kpc$^{-2}$, with a mean value of 28.8 \Msun yr$^{-1}$ kpc$^{-2}$. Regarding the control sample, it shows a comparable range of values but a lower mean value of 7.6 \Msun yr$^{-1}$ kpc$^{-2}$.

\subsection{Comparison with NIRSpec spectra}\label{sec:jwst_nirspec}

A total of 39 EELG candidates have NIRSpec spectra (see Table \ref{tab:nirspec}). Two galaxies have spectra but in one of them (CEERS48859), the [OIII] would fall in the detector gap according to the $z_{phot}$, and no line measurements are performed in this galaxy. In the other galaxy (CEERS26436) [OIII] also falls in a detector gap but [OII]$\lambda\lambda$3727,3729 (hereafter [OII]) and H$\alpha$ are measured in this galaxy. From the sample of EELG candidates with NIRSpec spectra, 32 of them have prism spectra while 15 of them have medium-resolution grating spectra. Here, we use the data products of the collaboration DR as described in Sec. \ref{sec:spectro_CEERS}. 
We first estimate the spectroscopic redshift with [OIII] using LiMe \citep[][for the technical details]{Fernandez2023,LIME2024}. We also measure the fluxes and EWs of H$\alpha$, and [OII]\footnote{The measurements are publicly available in \url{https://ceers-data.streamlit.app/}}. We report the obtained fluxes in Table \ref{tab:nirspec}. Regarding the control sample, there are 44 galaxies with NIRSpec spectra. [OIII] is detected in 27 of those spectra, 17 with the prism and 10 with the medium-resolution gratings. In the other 17 galaxies (38\%) with spectra, no bright emission lines are detected or the lines fall in detector gaps according to their $z_{phot}$.

\begin{figure}[!t]
    \centering
    \includegraphics[width=0.49\textwidth]{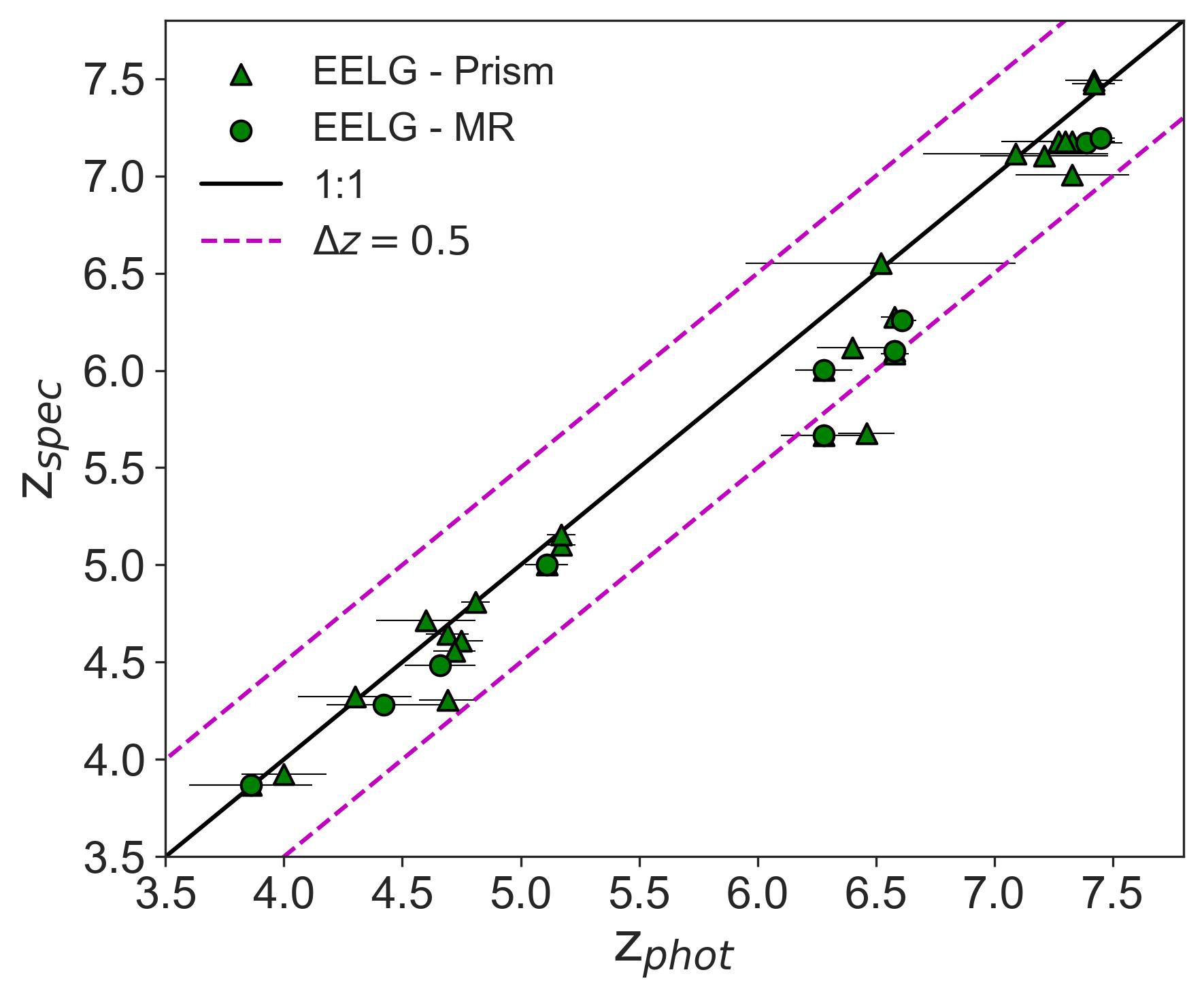}
    \caption{Comparison between photometric and spectroscopic parameters for the subsample of EELG candidates with NIRSpec spectrum. Photometric vs spectroscopic redshift with medium resolution (MR) grating (circles) and prism (triangles). The magenta dashed line is a difference of 0.5 in redshift. The black solid line is the 1:1 relation.}
    \label{fig:comparison_spectrum_z}
\end{figure}

In Fig. \ref{fig:comparison_spectrum_z}, we compare the photometric and the spectroscopic redshifts of the sample of EELGs. We find that photometric redshifts are on average $\Delta z = 0.2$ higher than the spectroscopic redshifts (and up to $\Delta z = 0.8$ higher).  The larger differences are in the cases where the photo-$z$ code confuses a galaxy with H$\alpha$ falling in the red edge of filter F444W with a galaxy with H$\alpha$ falling in the red edge of filter F410M or [OIII] falling in the blue edge of F356W. These degeneracies are due to the fact that strong emission lines fall on the edge of observed filters producing the largest differences in redshift in our sample.

\begin{figure}[!t]
    \centering
    \includegraphics[width=0.49\textwidth]{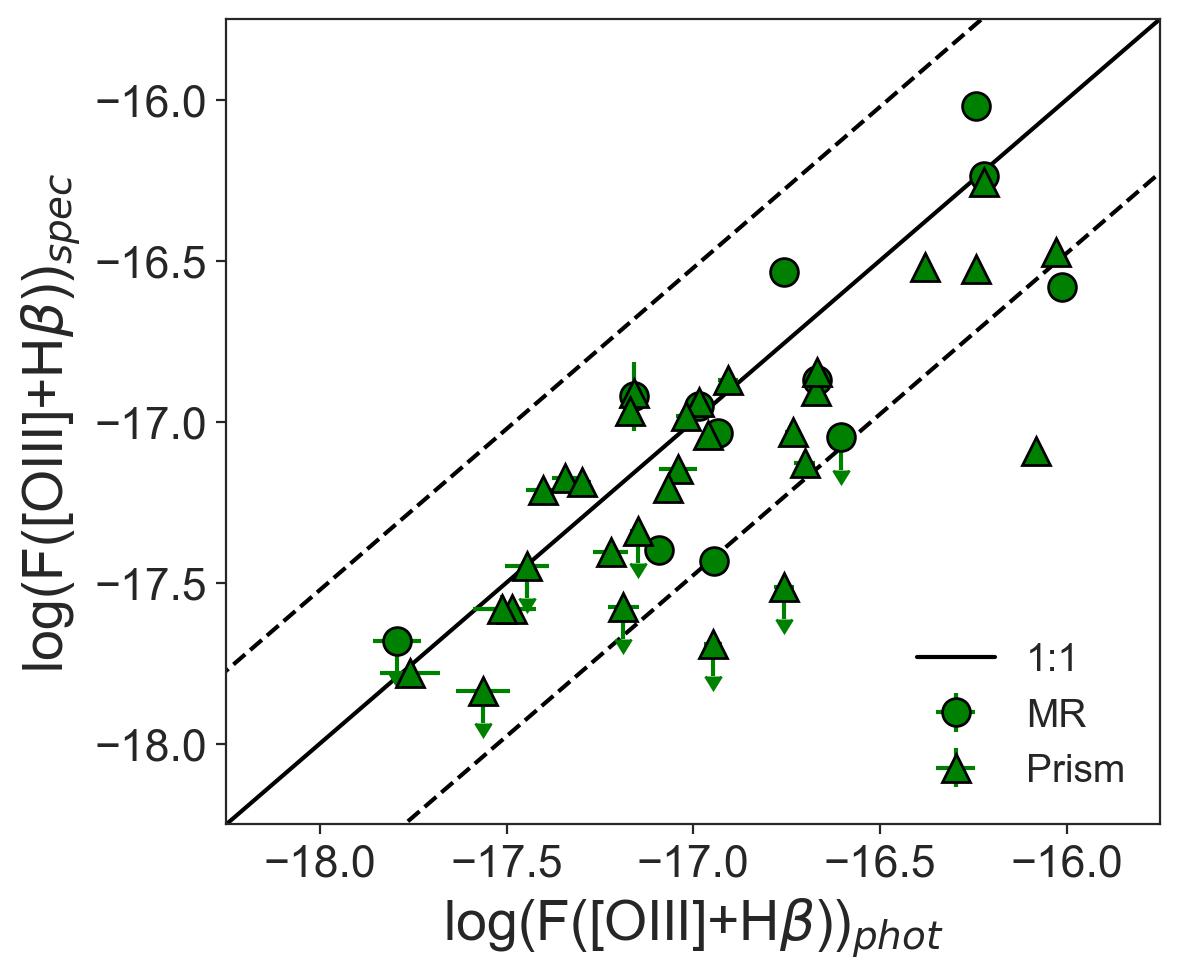}\\
    \includegraphics[width=0.49\textwidth]{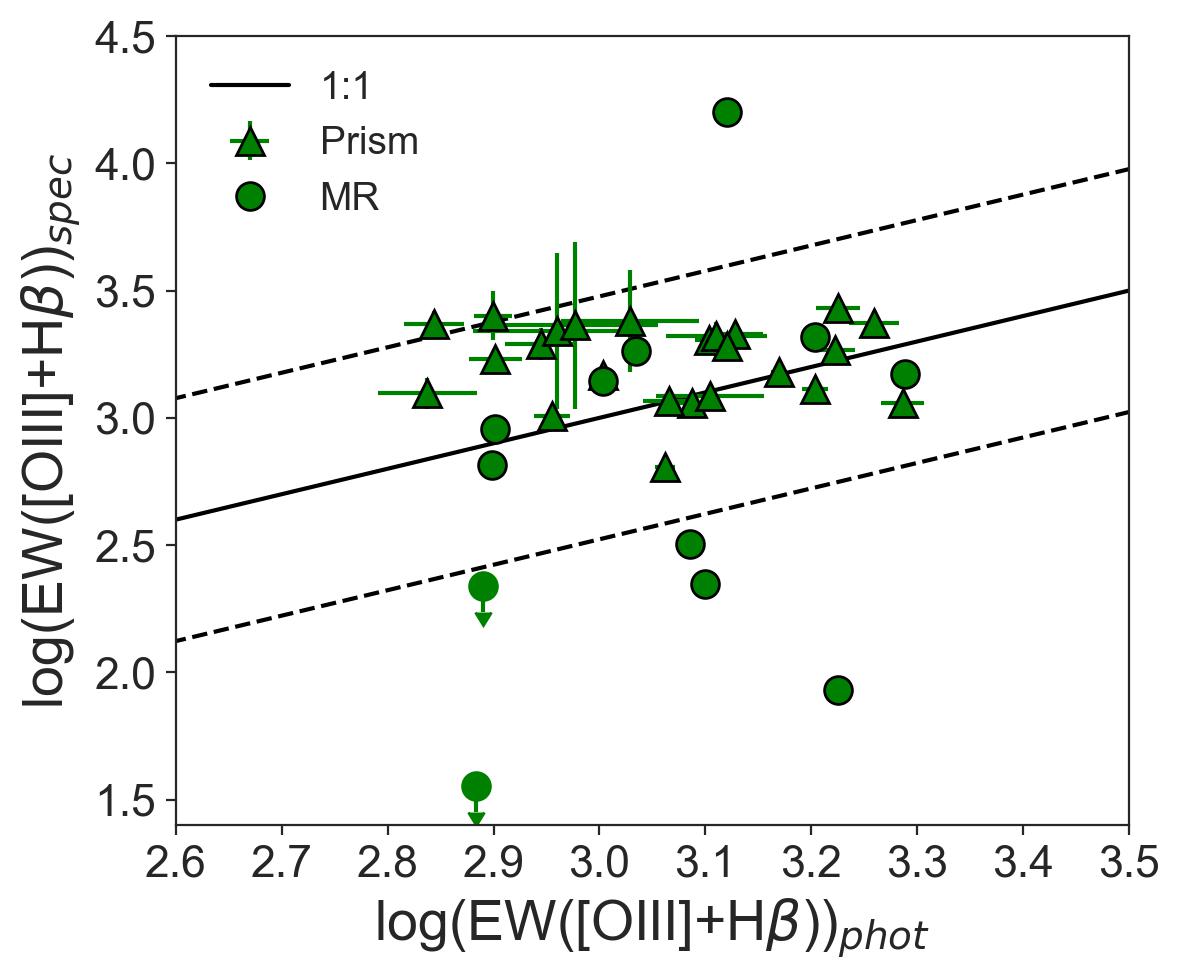}
    \caption{Comparison between photometric and spectroscopic parameters for the subsample of EELG candidates with NIRSpec spectrum. \textit{Top (Bottom) panel:} [OIII]+H$\beta$ flux ([OIII]+H$\beta$ EW) based on photometry vs spectra with grating (circles) and prism (triangles). 
    In both panels, the black dashed lines are a difference by a factor of 3. The black solid line is the 1:1 relation.}
    \label{fig:comparison_spectrum}
\end{figure}

We also compare the line flux [OIII]+H$\beta$ from the spectra and the photometric method (see the top panel in Fig. \ref{fig:comparison_spectrum}). The prism spectra show line flux within a factor of 0.1 and 1.77, with a mean value of 0.9, and most of the galaxies are within a factor of 3, compared to the photometric-based fluxes. Similar values are found in the medium-resolution spectra where the mean difference is a factor of 0.96, with values ranging from a factor of 0.27 to 1.72. We perform a K-S test and we find that the differences are significant between the spectroscopic fluxes and the ones estimated in Sec. \ref{sec:photo-fluxes}. The disagreement could be due to a mismatch of the photometric calibration or residual slit loss correction. But the fact they have [OIII] in emission indicates that the main contributor to the flux excess in the observed fluxes is in fact dominated by strong emission lines. We also compare the EW([OIII]+H$\beta$) from photometry and the spectra (see the bottom panel in Fig. \ref{fig:comparison_spectrum}) and we find they also agree within a factor of 3, especially in the cases where the spectra are taking from the prism configuration. In the cases with MR spectra, the differences may be larger due to the lack of detecting continuum which makes it more uncertain to quantify the EW of a given line.{ We omitted the errorbars for data points from MR spectra (green circles) because they are huge and unrealistic given that the continuum is not detected.}

In order to probe the mean properties of the sample, we performed two stacked spectra. One includes all galaxies with prism spectra and the other includes the galaxies with medium-resolution spectra. To perform the stacking, we calculate the median of all spectra in each bin of rest-frame wavelength. The results of this stacked analysis are presented in the following sections along with measurements in individual galaxies.

\section{Results and discussion}\label{sec:results-discussion}

\subsection{Young starbursts}\label{sec:jwst_bpt}

\begin{figure}[!t]
    \centering
    \includegraphics[width=0.5\textwidth]{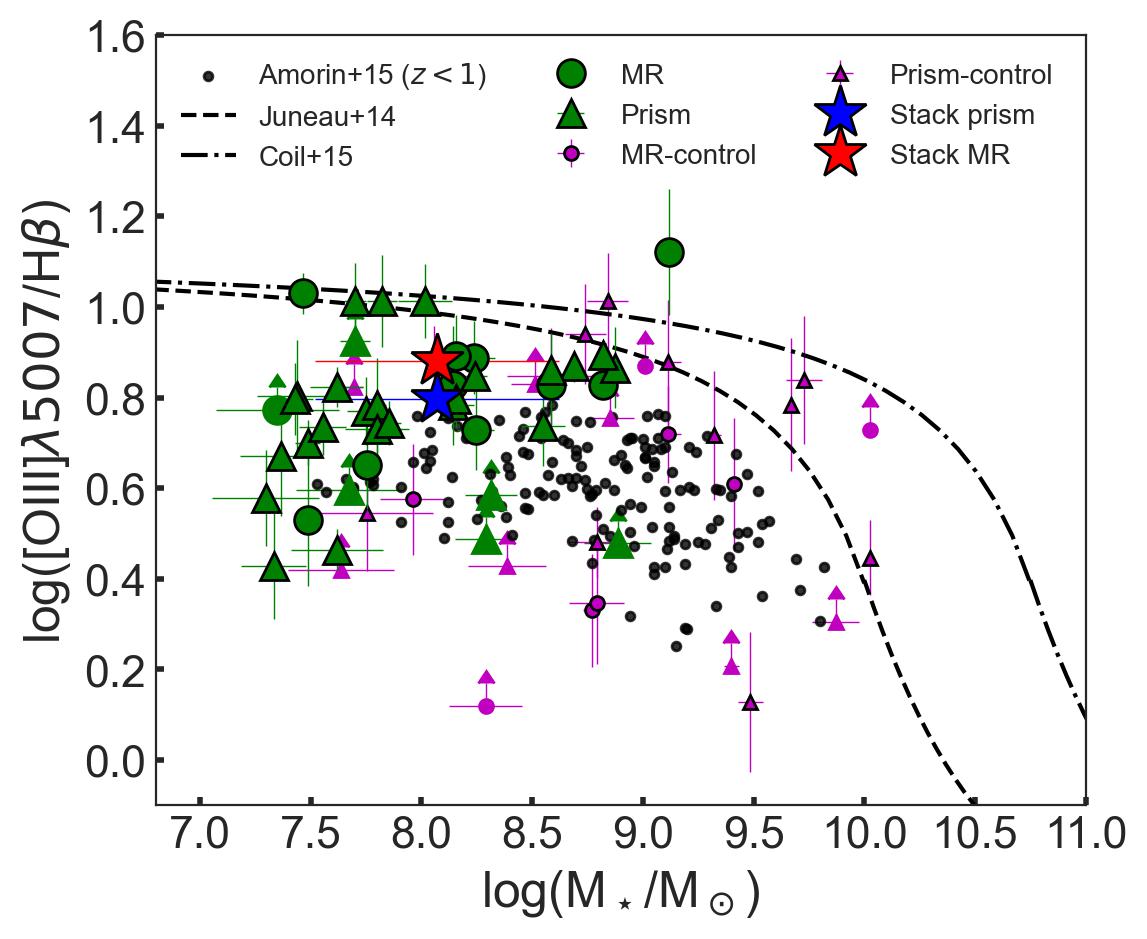}
    \caption{Mass-Excitation (MEx) Diagram. The EELG candidates with NIRSpec spectra are in green symbols (circles for medium-resolution (MR) and triangles for prism). The stacks in this work are marked by star symbols. Galaxies in the control sample with NIRSpec spectra are shown with magenta circles (medium-resolution mode) and triangles (prism mode) symbols. The black dashed line indicates the demarcation between star-forming galaxies and AGN, according to \cite{Juneau2014}. The black dashed-dotted line is the demarcation at $z\sim 2$ \citep{Coil2015}. The black circles are EELGs up to $z\sim1$ \citep{Amorin2015}. }
    \label{fig:Mex}
\end{figure}

We use the Mass-Excitation \citep[MEx, ][]{Juneau2014} diagram to study the ionization source in the sample of EELGs. We note that MEx is usually not preferred to other line ratio diagnostics given the added biases of using stellar masses \citep[e.g.][]{Coil2015,Cleri2023clear}, but we keep our analysis based on the MEx diagnostic because we can apply it to a large amount of galaxies in the sample. Since we can not resolve [OIII] and H$\beta$ from the photometry, we focus our analysis on the subsample with spectroscopy. In Fig. \ref{fig:Mex}, we show the position of our sample of EELGs in the classical MEx diagram. We find that their position is consistent with the location of metal-poor starbursts with high log([OIII]$\lambda$5007/H$\beta)\gtrsim 0.4-1$. This region is populated by local analogs of these high-$z$ systems. For example, in Fig. \ref{fig:Mex}, we include the sample of EELGs up to $z\sim 1$ from \cite{Amorin2015} to compare with our sample. We notice that the sample of low-$z$ EELGs are in the same locus with slightly higher stellar masses.  We also show that the stacked spectra show similar [OIII]/H$\beta\sim 7$ ratios in both cases using prism or medium-resolution. This indicates the high ionization conditions in this subsample of EELGs. Regarding the control sample with confirmed spectra, we note in Fig. \ref{fig:Mex} that they show similar values of log([OIII]/H$\beta$), even though there are more galaxies where H$\beta$ is an upper limit. 

The excitation properties of low-mass compact systems are consistent with being dominated by young starbursts. However, we cannot rule out other non-thermal sources (e.g., AGN) for the entire sample of photometric-selected EELGs using only photometry \citep[e.g.][]{Backhaus2023,Barro2023,Cleri2023,Davis2023,Larson2023BH}.

\begin{figure}[!t]
    \centering
\includegraphics[width=0.49\textwidth]{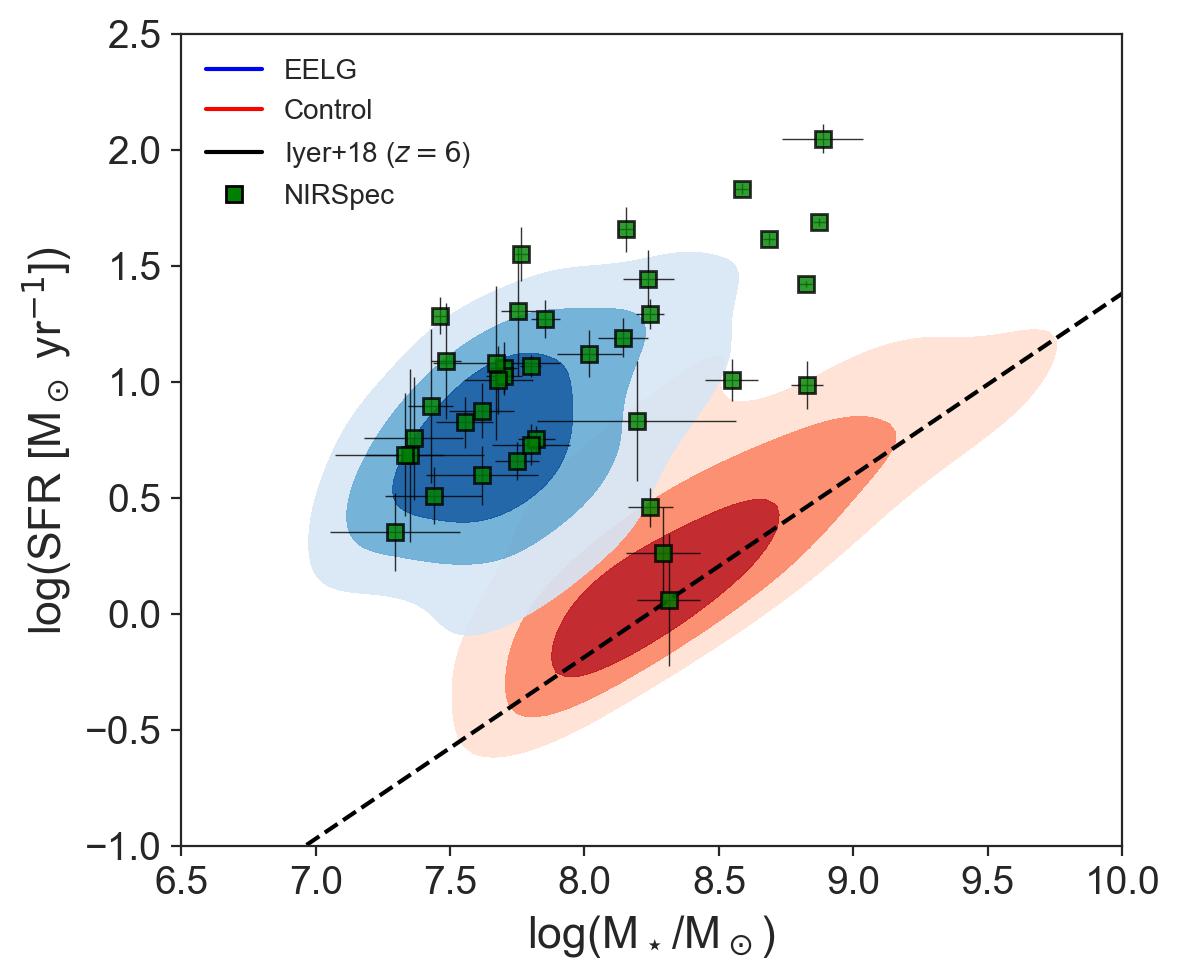}
    \caption{SFR - M$_{\star}$ diagram for our sample of EELGs (in blue scale contours) and control sample (in red scale contours). The contours are the percentiles 75th, 50th, and 25th respectively. The solid line is the main sequence from \cite{Iyer2018} at $z\sim 6$ that covers a mass range from 10$^6$\Msun. The green squares are the subsample of EELGs with NIRSpec spectra.}
    \label{fig:MS_young}
\end{figure}

In Fig. \ref{fig:MS_young} we show the position of the sample of EELGs and the control sample in the stellar mass-SFR plane. The contours are the 75th, 50th, and 25th percentiles, respectively. We find that the galaxies in the control sample are distributed along the so-called main sequence \citep[e.g.][]{Speagle_2014} with smaller scatter compared to the observed scatter in the EELGs candidates. The sample of EELGs shows larger SFR for a given stellar mass compared with the control sample, which implies higher sSFR, as previously shown in Fig. \ref{fig:SED_parameters_EELGs}. Some galaxies may be part of the main-sequence but most of the sample shows higher SFR which implies they are likely strong starbursts. {This bimodality between main-sequence galaxies and starbursts has been observed in other works at $z\sim 3-6.5$ \citep[e.g.][]{Rinaldi2022}.} Given that our sample covers a wide range of redshifts, we assume no evolution for $z>3$ in the normalization of the main-sequence but an increase of the scatter at $z>4$ due to more bursty SFH \citep[e.g.][]{Cole2023}. Individual galaxies with NIRSpec spectra are also shown in Fig. \ref{fig:MS_young} and they show a wide range in stellar mass ($\sim$2 dex) and in SFR ($\sim$2 dex). We note that given the wide range of parameters, this subsample is representative of all galaxies in the entire sample as will also be seen in the following Fig. \ref{fig:ssfr-age}.

EELGs also show systematically higher sSFR than the control sample, as can be seen in Fig. \ref{fig:ssfr-age} (top panel). We also found an increase of sSFR with EW([OIII]+H$\beta$). This increase starts for sSFR$>10^{-8}$ yr. This trend is not observed in the control sample where the scatter is larger for a given EW. Similar results are found with the age of the starburst (see the bottom panel in Fig. \ref{fig:ssfr-age}) where the youngest bursts show the largest EW([OIII]+H$\beta$). We compare these results with a sample of [OIII] emitters at $z\sim1.3-2.4$ \citep{Tang2019} and we find our results are consistent with the trend observed of intermediate-$z$ analogs. A similar trend of EW with sSFR is found in \cite{Papovich2022} for $\sim$ 200 star-forming galaxies at $z\sim 1.1-2.3$ using HST/WFC3 IR grisms. {Similarly, the decreasing trend of age with EWs has also been found in other works of emission-line galaxies at $z\sim7-8$ \cite{Rinaldi2023}.}

\begin{figure}[!t]
    \centering
\includegraphics[width=0.5\textwidth]{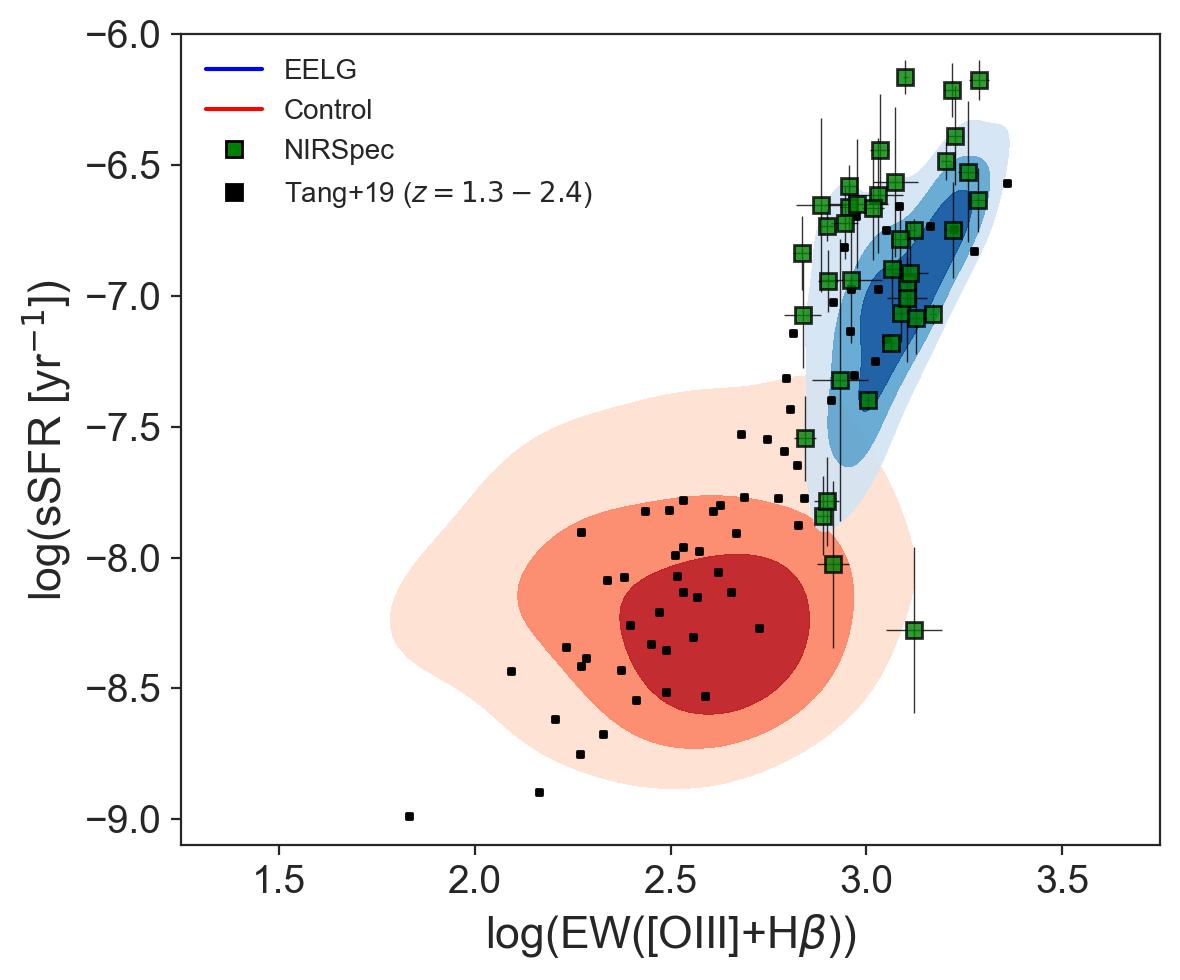}\\\includegraphics[width=0.5\textwidth]{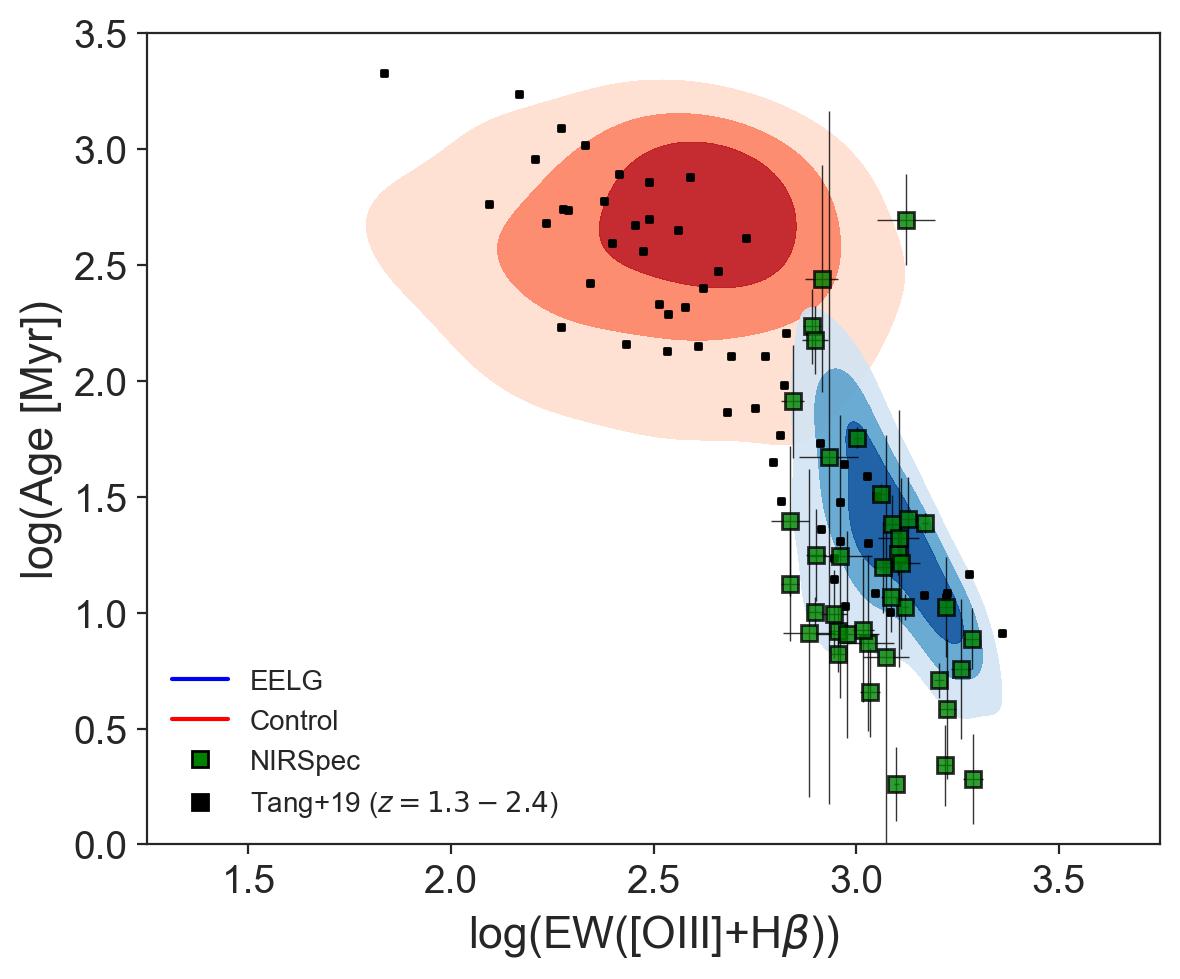}
    \caption{Relation between EW([OIII]+H$\beta$) and sSFR (top panel) and age (bottom panel) for the sample of EELGs and control sample (same as in Fig. \ref{fig:MS_young}). The black squares are a sample of EELGs at $z=1.3-2.4$ \citep{Tang2019}. The EWs of the sample with NIRSpec spectra are from photometry measurements.}
    \label{fig:ssfr-age}
\end{figure}

\subsection{Ionising photon production efficiency $\xi_{\rm ion}$}\label{sec:scaling}

Ionizing photon production efficiency $\xi_{\rm ion}$ refers to the ratio of ionizing photons emitted by stars within a galaxy to the total number of photons produced by those stars. This efficiency is a crucial parameter to estimate the contribution of different types of galaxies to the reionization process. Galaxies with high ionizing photon production efficiencies are likely to play a more significant role in driving reionization. Scaling relations have been proposed to estimate such parameters based on observables, for example, the $\xi_{\rm ion}$-EW([OIII]) relation proposed in \cite{Chevallard2018}.

\begin{figure}[!t]
    \centering
\includegraphics[width=0.5\textwidth]{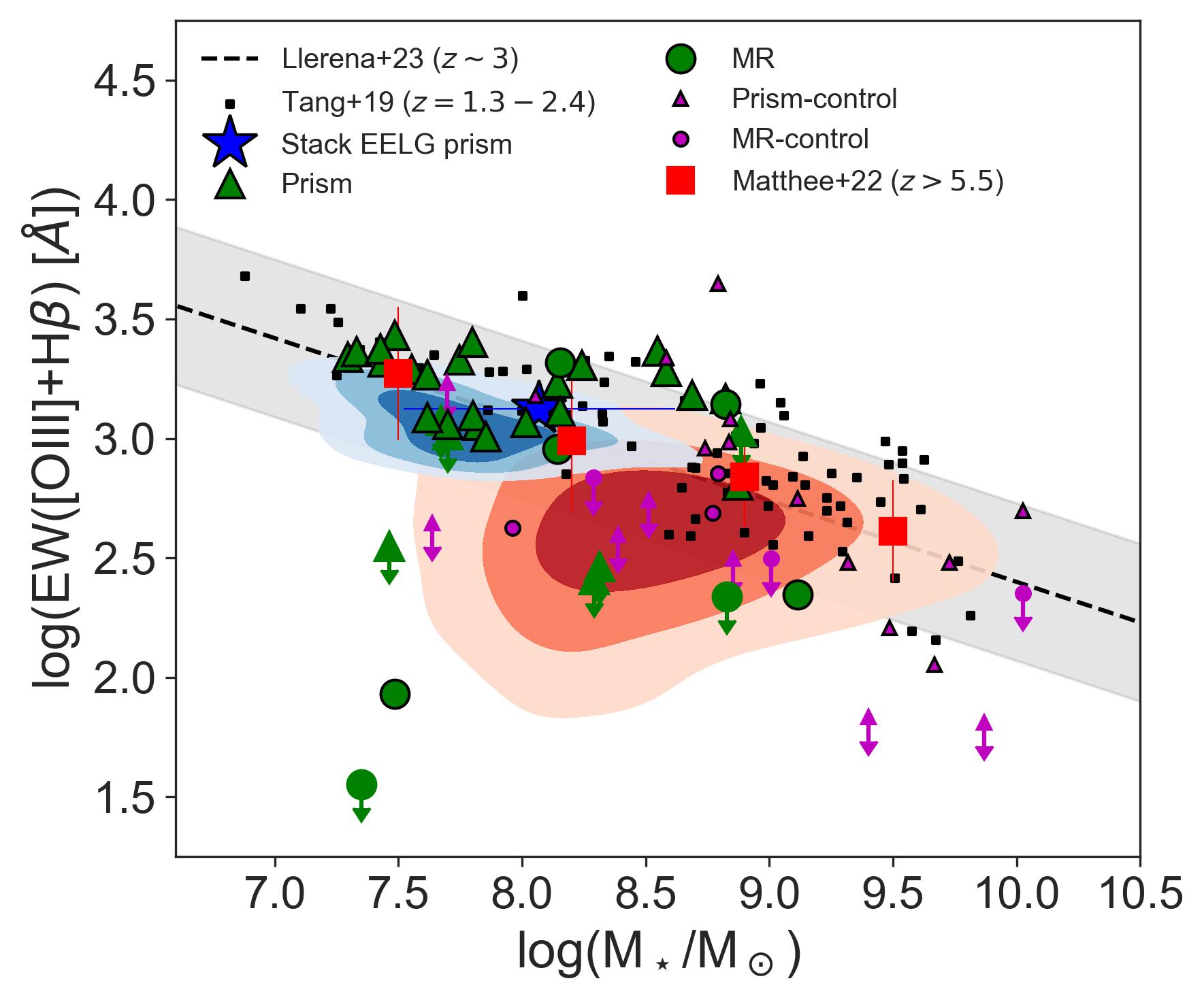}
    \caption{Relation between stellar mass and EW([OIII]+H$\beta$). The contours are the same as in Fig. \ref{fig:MS_young}. In green symbols, the subsample of EELG candidates with NIRSpec spectroscopy, while the control sample is in magenta symbols.  The blue star is the stack of the EELGs with prism spectra. The black dashed line is the relation at $z\sim 3$ and the observed scatter \citep{Llerena2023}. The red squares are stacks of $z>5.5$ galaxies \citep{Matthee2023}.}
    \label{fig:EW-mass-EELG}
\end{figure}

In Fig. \ref{fig:EW-mass-EELG}, we show the scaling relations between EW([OIII]+H$\beta$) and the stellar mass. A clear negative correlation between emission line EW and stellar mass has been found in the literature at different redshifts\citep[e.g.][]{Tang2019,Lumbreras-Calle2022,Matthee2023}. This indicates that galaxies with lower masses tend to have stronger recent star formation events relative to their mass (higher sSFR). However, we note that given the high average value of [OIII]/H$\beta$ we measured, then EW([OIII]+H$\beta$) is dominated by [OIII], and the strength of [OIII] can depend on other excitation effects (e.g., metallicity and the hardness of the ionizing spectrum), and not only on stellar mass or sSFR. As shown in Fig. \ref{fig:EW-mass-EELG}, we find that the spectroscopically-confirmed EELGs in our sample follow the trend observed at lower redshift ($z\sim$ 3, black dashed line in Fig. \ref{fig:EW-mass-EELG}). This trend seems to be followed also by spectroscopically-confirmed galaxies in the control sample, with few cases with higher stellar mass and an overlap region with the EELGs region. This is consistent with other works based on stacking of $z>5.5$ galaxies \citep[red squares, ][]{Matthee2023}. Overall, galaxies with low stellar masses and high sSFR tend to show higher EWs and then are more likely to show extreme ISM conditions reflected in their extreme EWs.
\begin{figure}[!t]
    \centering
\includegraphics[width=0.5\textwidth]{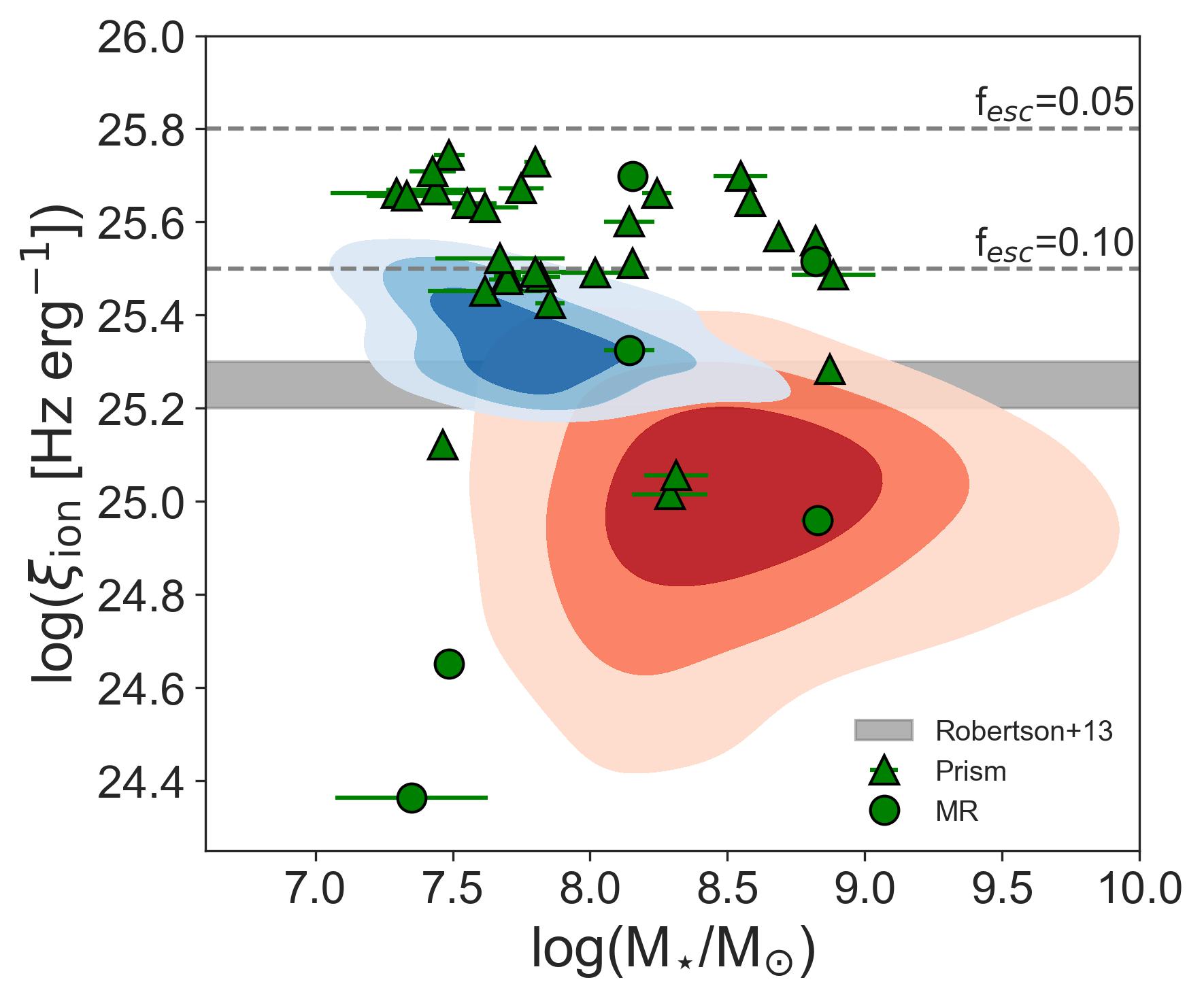}
    \caption{Relation between stellar mass and $\xi_{\rm ion}$ for the sample of EELGs and control sample (same as in Fig. \ref{fig:MS_young}). The gray shaded region is the canonical $\xi_{\rm ion}$ value given by a simple stellar population at constant SFR over 100Myr \citep{Robertson2013}, assuming an escape fraction of 0.20. The horizontal dashed lines are the canonical $\xi_{\rm ion}$ values for an escape fraction of 0.10 and 0.05, respectively.}
    \label{fig:xi-ew}
\end{figure}

Based on  photometry, our EELGs show a mean EW([OIII]+H$\beta$)=1231\r{A} ($\sigma=421$\r{A}) with values as high as 2932\r{A}. The stack from prism spectra shows a typical EW([OIII]+H$\beta$) of 1258\r{A}. The highest value measured in a single galaxy with NIRSpec/prism spectra is EW([OIII]+H$\beta$)=2553\r{A}. For H$\alpha$, the mean EW is 808\r{A} ($\sigma=392$\r{A}), which is consistent with the EW(H$\alpha$)=793\r{A} of the stack. 
We use the relation in \cite{Tang2019} depending on EW([OIII]$\lambda$5007) and we extrapolate the relation to cover the EWs observed in the control sample and in the EELG sample. This relation is consistent with recent observations of galaxies at $z=3-5.7$ \citep{Boyett2024}. We obtain $\xi_{\rm ion}$ for both samples. The values are displayed in Fig. \ref{fig:xi-ew}. We find that roughly 65\% of the EELGs (483 out of 736 candidates) show $\xi_{\rm ion}$ higher than the canonical values for a simple stellar population at constant SFR over 100Myr \citep{Robertson2013}. This subsample of EELGs, in particular the galaxies with lower stellar masses (due to the mild trend we find), is efficiently producing ionizing photons and are ideal laboratories to probe the escape of ionizing photons. Actually, our sample may include strong candidates of reionization galaxies with LyC escape as will be discussed in Sec. \ref{sec:LyC_escape}. As expected from the lower EWs of the control sample, they show $\xi_{\rm ion}$ below the canonical values, and only 3\% of the control sample show values within the range of canonical values. The sample of EELGs shows a mean value of $\log\xi_{\rm ion} [{\rm Hz/erg}]=25.35$ while the control sample shows a mean value of $\log\xi_{\rm ion} [{\rm Hz/erg}]=24.90$.

\subsection{Gas-phase metallicity}\label{sec:metallicity_gas}

\begin{figure}[!t]
    \centering
\includegraphics[width=0.5\textwidth]{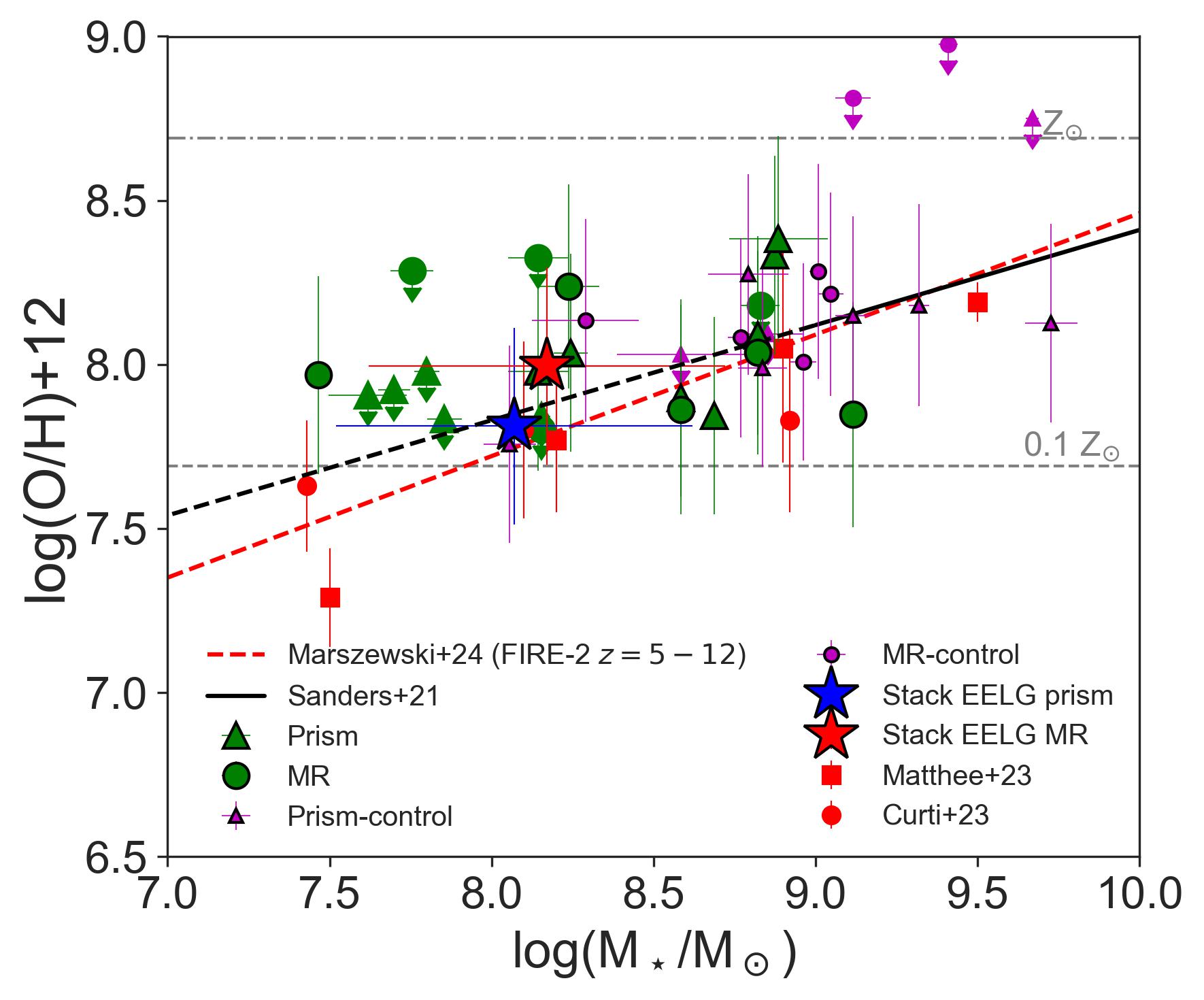}
    \caption{Mass-Metallicity relation for EELGs. Individual galaxies of the EELG sample (control sample) are in green (magenta) symbols. Results from the stacks are represented with red (grating) and blue (prism) stars. Red squares are stacks at $z\sim 5-7$ from \cite{Matthee2023} and red circles are stacks at $z\sim 3-10$ from \cite{Curti2023}. The black solid line is the MZR at $z\sim3.3$ from \cite{Sanders2021} and its extrapolation to lower stellar masses in the dashed line. {The red dashed line is the predicted MZR at $z=5-12$ from FIRE-2 simulation \citep{Marszewski2024}.} }
    \label{fig:MZR}
\end{figure}
For the galaxies (20 out of 39) with a detected H$\alpha$ and H$\beta$ ({S/N$>3$}), we estimated the nebular attenuation assuming H$\alpha$/H$\beta=2.79$ under case B approximation for T$_{\rm e}=15000$K and $n_{\rm e}=100$ cm$^{-3}$ \citep{EPM2017} and considering the Cardelli law \citep{Cardelli1989}. For galaxies with prism, we find a mean value of 0.29 mag (H$\alpha$/H$\beta$=3.79), while for medium-resolution we find a mean value of 0.21 mag (H$\alpha$/H$\beta$=3.31). We use these mean values to estimate the nebular attenuation in galaxies with unobserved H$\alpha$ or undetected H$\beta$. We corrected the observed fluxes reported in Table \ref{tab:nirspec} assuming the \cite{Reddy2015} law with $R_V=2.505$.

We used the O32=[OIII]$\lambda5007$/[OII]$\lambda\lambda$3727,29 calibration from \cite{Bian2018} to estimate the gas-phase metallicity in individual galaxies and in the stacks. In Fig. \ref{fig:MZR}, we show the gas-phase metallicity for the sample of EELGs with NIRSpec spectra. We estimate gas-metallicity only in the subsample of galaxies where [OII] is detected or where we were able to put an upper limit. The galaxies with detected [OII] in prism spectra show a mean redshift of $z\sim 5.39$ and a mean gas-phase metallicity of $\log$(O/H)+12=8.07 ($\sim 24$\% solar). On the other hand, the galaxies with detected [OII] in medium-resolution spectra show a mean redshift of $z\sim 5.96$ and a mean gas-phase metallicity of $\log$(O/H)+12=8.00 ($\sim 20$\% solar). Regarding the stacks, we find consistent values of $\log$(O/H)+12=7.85 ($\pm$0.3) and 8.01 ($\pm$0.3) for prism and grating stacks, respectively. These values are not different from the typical average metallicities measured in EELG at low and intermediate redshifts \citep[e.g.][]{Amorin2010,Amorin2014,Maseda2014,Calabro2017,Perez-Montero2021,Tang2021}. {We also note that the obtained values are consistent with the gas-phases metallicities of the templates ASK 17, 20, 21 with values 0.25-0.35Z$_{\odot}$}. We also estimate the metallicities using the calibration for EELGs presented in \cite{Perez-Montero2021} and we find a mean difference of 0.03 towards lower metallicities using this O32 calibration, which means no significant differences are using both calibrations. We also explore the same method used in \cite{Sanders2021} based on multiple diagnostics to estimate the metallicity minimizing their Eq. 6. We used the diagnostics [OIII]$\lambda$5007/H$\beta$, [OII]$\lambda$3727/H$\beta$, O32 and R23=$\dfrac{[OIII]\lambda\lambda4959,5007+[OII]\lambda3727}{H\beta}$. We find a mean difference of 0.1 dex compared with using only the O32 calibration from \cite{Bian2018}, which would reduce our metallicities to $\sim$15-20\%. Similar differences of up to 0.1 dex are found using the low metallicity branch of the R23 calibration from \cite{Papovich2022} which is tested in HST grism spectra with galaxies at $z\sim 1.1-2.3$.

We note that these values are consistent with the mass-metallicity relation at $z\sim3.3$ reported in \cite{Sanders2021}, and {comparable with those predicted in FIRE-2 simulation at $z\sim5-12$ \citep{Marszewski2024}.} Compared with other recent works, our results are comparable to those of \cite{Matthee2023} at $z\sim 5-7$ and \cite{Curti2023} at $z\sim 3-10$ in the similar stellar mass range. The metallicities for the galaxies in the control sample tend to also follow the MZR as can be seen in Fig. \ref{fig:MZR}, with slightly higher values in the high mass end compared to the EELG sample.

Overall, we find that our sample includes galaxies with subsolar metallicities at around $\sim$20-25\% solar, consistent with the mass-metallicity relation, which suggests they do not show different metallicities compared with the general population at similar redshifts.

\subsection{Conditions for the  escape of ionizing photons}\label{sec:LyC_escape}
\begin{figure}[!t]
    \centering
\includegraphics[width=0.5\textwidth]{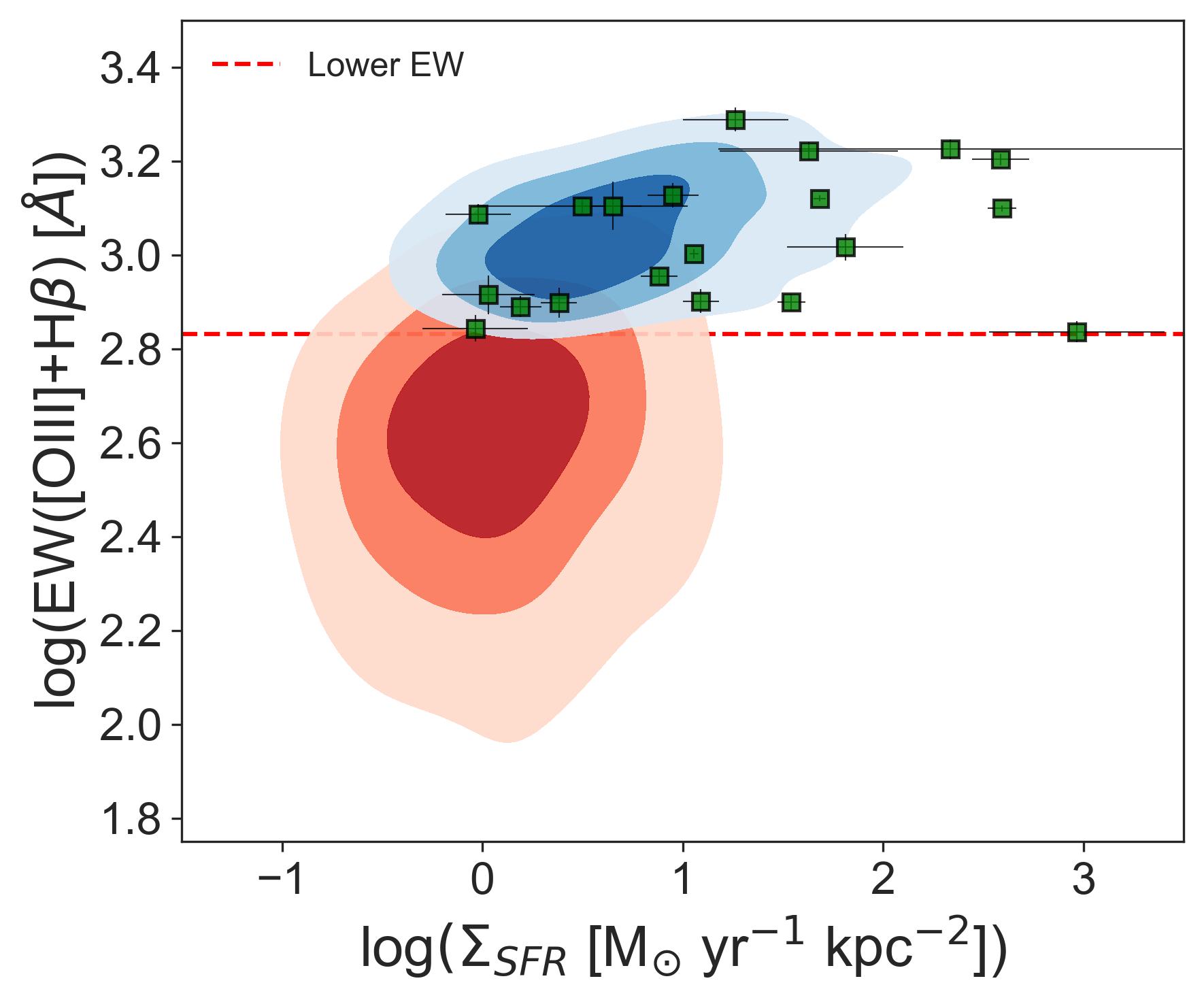}
    \caption{Relation between EW([OIII]+H$\beta$) and $\Sigma_{\rm SFR}$ for the sample of EELGs and control sample (same as in Fig. \ref{fig:MS_young}). The green squares are the subsample of EELGs with NIRSpec spectra. The red dashed line is the lower EW limit of the ASK templates in this paper.}
    \label{fig:sigma-ew}
\end{figure}

\begin{figure}[!t]
    \centering
\includegraphics[width=0.5\textwidth]{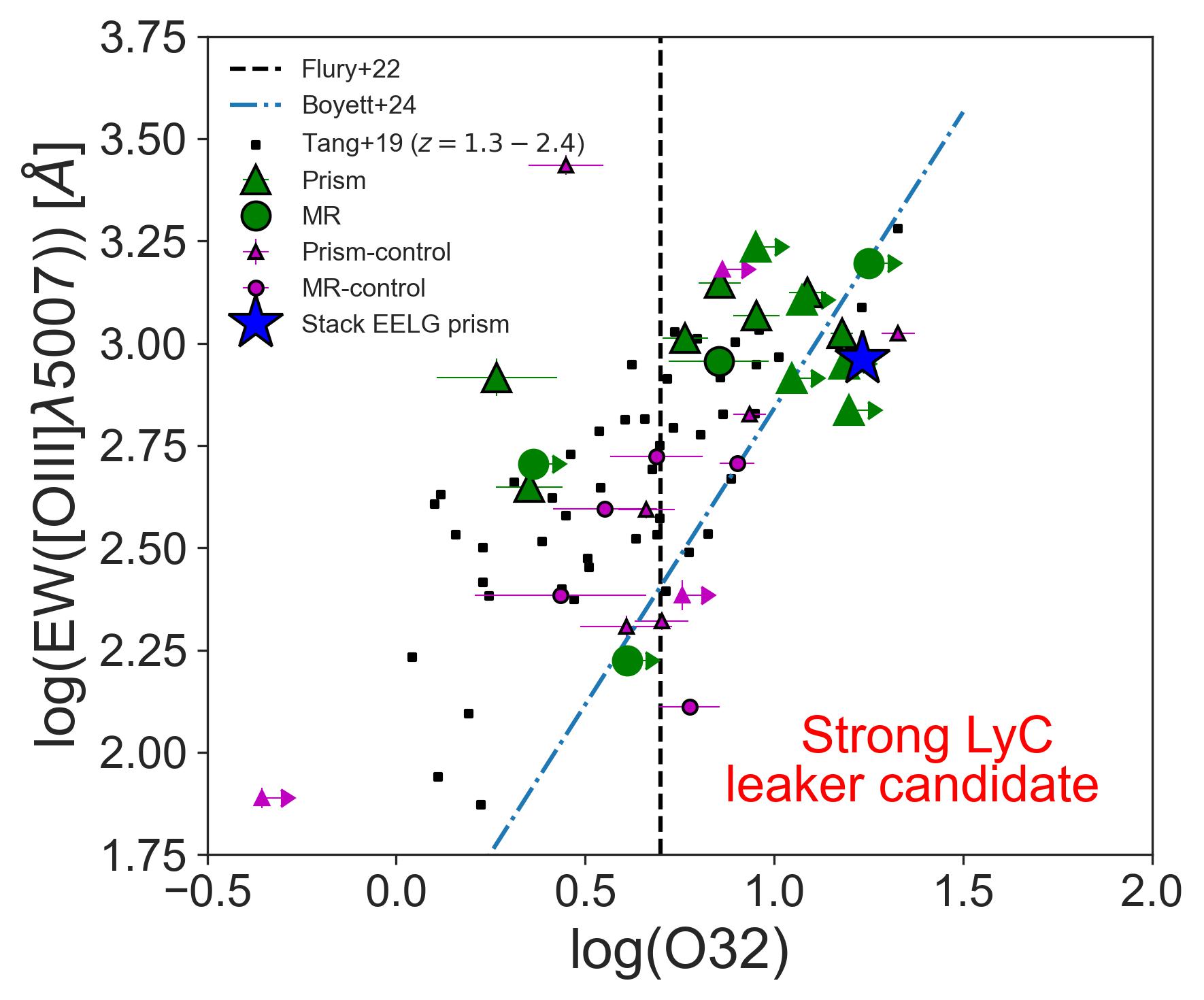}\\\includegraphics[width=0.5\textwidth]{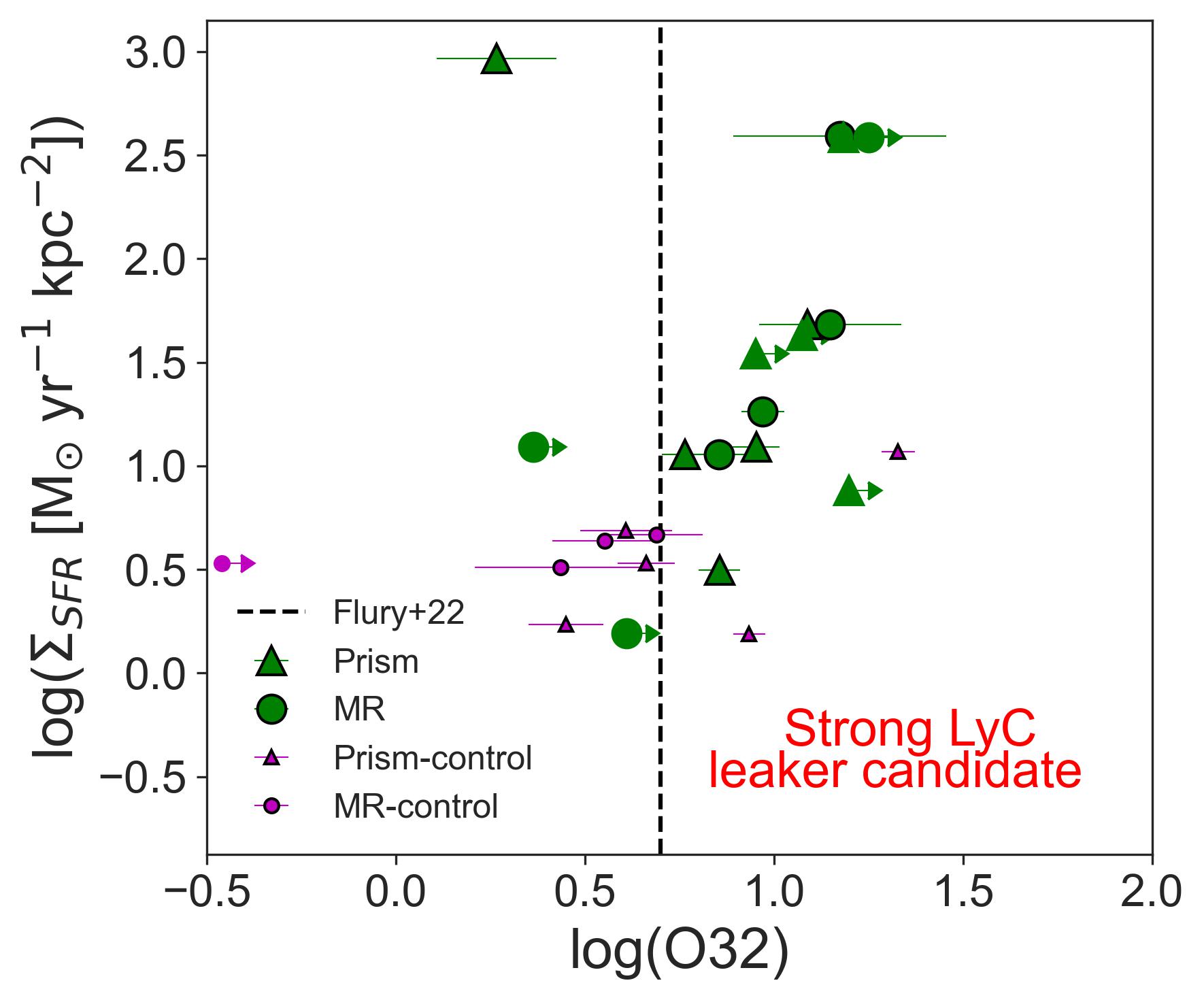}
    \caption{Relation between O32 and EW([OIII]$\lambda$5007) (top panel) and $\Sigma_{\rm SFR}$ (bottom panel). We include lower limits due to the non-detection of [OII]. In green symbols, the sample of EELGs with NIRSpec spectra, while in magenta symbols, the control sample with NIRSpec spectra. The vertical dashed line is the separation between weak and strong LyC leaker according to \cite{Flury2022}. On the top panel, black squares are a comparison sample at $z=1.3-2.4$ \citep{Tang2019}, and the blue dashed-dotted line is the relation found in \cite{Boyett2024} for a sample of galaxies at $z=3-9.5$.}
    \label{fig:o32-ew}
\end{figure}

In this section, we explore indirect tracers of the escape of ionizing photons to understand if our sample of young starburst is made of strong LyC leakers. We use indirect tracers since in galaxies in the EoR, as the ones included in this paper, a direct probe of the LyC is not possible due to the high fraction of neutral hydrogen and then indirect tracers are required \citep[e.g.][]{Mascia2023CEERS}.

We first explore the relationship with the star formation surface density \citep[e.g.][]{Flury2022}. As displayed in Fig. \ref{fig:sigma-ew}, we find that our EELG candidates tend to show higher EWs while increasing the $\Sigma_{\rm SFR}$. Regarding the spectroscopically-confirmed EELGs, we find that they show high $\Sigma_{\rm SFR}$ at high EW. Regarding the control sample, we show that they have $\Sigma_{\rm SFR}\lesssim 10$ \Msun yr$^{-1}$ kpc$^{-2}$, while the EELGs show up to larger values. We note that $\Sigma_{\rm SFR}\gtrsim 10$ \Msun yr$^{-1}$ kpc$^{-2}$ is the range in which the fraction of detected LyC leakers is $>40$\% according to observations of local galaxies \citep{Flury2022}. This indicates that galaxies with higher EWs and concentrated SFR have the conditions that should facilitate the escape of LyC photons in agreement with the f$_{\rm esc}\propto\Sigma_{\rm SFR}^{0.4}$ relation described in \cite{Naidu2020}.

We explore O32 as an alternative indicator of LyC leaking \citep[e.g.][]{Izotov2020}. In Fig. \ref{fig:o32-ew} (top panel), we show the relation of O32 with EW([OIII]$\lambda$5007). We show there is an increase of EW with O32. We find that 9 out of 12 (75\%) spectra with detected [OII] and [OIII] in their spectra show O32$>5$. The galaxies with EW([OIII]$\lambda$5007)$\gtrsim$ 400\r{A} are the ones that show O32$>5$ which indicates strong LyC leaker candidates based on local galaxies \citep{Flury2022}. We note that a value of O32$=5$ corresponds to a metallicity of $0.27$Z$_{\odot}$ using the \cite{Bian2018} calibration, which is approximately the mean metallicity of the sample of EELGs. 
The trend we find is consistent with analogs at lower redshifts \citep{Tang2019}. The stack with galaxies with prism spectra show O32=14.6 and EW([OIII])=874\r{A} (blue star in Fig. \ref{fig:o32-ew}), which indicates that is a population of galaxies with conditions of escape of ionizing photons and deeper observations are needed in individual galaxies to probe their nature. Regarding the control sample, we note a lower fraction of strong candidates to be strong leakers.

As expected from previous results, we find an increase of $\Sigma_{\rm SFR}$ with O32 (bottom panel in Fig. \ref{fig:o32-ew}). This indicates that indeed the EELGs with higher concentrations of star formation provide the feedback necessary to clear LyC escape paths in the ISM. Similarly, we note a lower fraction of galaxies in the control sample with a high concentration of star formation compared to the sample of EELGs. We find also a lower fraction of strong candidates of strong leakers in the control sample.

\section{Conclusions}\label{sec:jwst-summary}

We use NIRCam photometry and empirical templates to select EELGs.  We demonstrate that NIRCam can identify a large sample of previously unknown EELGs in a wide range of redshifts showing unique properties similar to the rare metal-poor local starburst. The proposed color selections can effectively identify galaxies with EW([OIII] + H$\beta$) $>680$ \r{A} at all redshifts targeted in this work. We use the broad-band filters F277W, F356W, and F444W  and the medium-band filter F410M to select EELGs at $4\lesssim z\leq 9$. We selected a sample of 1000 candidates. 47 of them have already NIRSpec spectra. We define a control sample to compare the properties with the sample of EELGs. We use BAGPIPES to estimate their physical properties considering a delayed $\tau$-model for the SFH and a nebular component to model the intense emission lines. We also use the F200W filter to estimate the physical size of the young stellar populations. We find:

\begin{itemize}
    \item Our sample of EELGs shows a mean stellar mass of $10^{7.84}$\Msun with high sSFRs with a mean value of $10^{-7.03}$ yr$^{-1}$. They are young with a mean value of the time after the onset of star formation of 45Myr. 
    \item Compared with the control sample, they show slightly lower stellar masses but similar stellar metallicities and dust attenuation. The larger differences are in the sSFR, ages, and ionization parameter. This suggests they may have similar underlying stellar populations but the young starburst may be the reason for the extreme emission lines. 
    \item Based on their confirmed emission lines, we find that they are in the locus of metal-poor starbursts with high log([OIII]/H$\beta$) $\gtrsim$0.4-1 which indicates that star-formation may be the dominant source of ionization in these galaxies. The stacking analysis confirms this result with a small sample of NIRSpec spectra. The starburst nature of these galaxies is also observed in their position above the main-sequence. 
    \item Based on the photometric fluxes, we find a mean rest-frame EW([OIII]+H$\beta$) of 1231\r{A} for our sample of EELGs. The EW([OIII]+H$\beta$) shows an increase with sSFR and a decrease with age and stellar mass. In the control sample, the scatter is larger in these relations. 
    \item We use the EWs to estimate the ionizing photon production efficiency and we found that roughly 65\% (483 out of 736 candidates) of the sample of EELGs show $\xi_{\rm ion}$ values higher than the canonical values, which implies they are efficiently producing ionizing photons and are ideal laboratories to probe the escape of ionizing photons. The sample of EELGs shows a mean value of $\log\xi_{\rm ion} [{\rm Hz/erg}]=25.35$. 
    \item We find sub-solar gas-phase metallicities for the sample of EELGs based on the O32 calibration with a mean value of 20-25\% solar. They follow the MZR at $z\sim3.3$ which suggests they do not show different gas-phase metallicities than the general population of galaxies.
    \item We find an increase of EW with $\Sigma_{\rm SFR}$. The sample of EELGs can reach $\Sigma_{\rm SFR}>$ 10 \Msun yr$^{-1}$ kpc$^{-2}$ which indicate they are strong candidates of LyC leakers. Another indirect indicator is the high values of O32$>$5 (corresponding to gas-phase metallicities $<0.27$Z$_{\odot}$) that can be reached for some galaxies in the sample. This indicates that they may have the conditions to facilitate the escape of ionizing photons.
\end{itemize}

\begin{acknowledgements}    
    {We thank the anonymous referee for the detailed review and useful suggestions that helped to improve this paper}. We wish to thank all our colleagues in the CEERS collaboration for their hard work and valuable contributions to this project. We wish to thank to Jorge Sánchez-Almeida for providing us with the ASK templates used in this paper. MLl acknowledges support from the PRIN 2022 MUR project 2022CB3PJ3 - First Light And Galaxy aSsembly (FLAGS) funded by the European Union – Next Generation EU. RA acknowledges the support of ANID FONDECYT Regular Grant 1202007. RA acknowledges financial support from the Severo Ochoa grant CEX2021-001131-S funded by MCIN/AEI/10.13039/501100011033.
This work has made extensive use of Python packages astropy \citep{astropy:2018}, numpy \citep{harris2020}, Matplotlib \citep{Hunter:2007} and LiMe \citep{LIME2024}.
\end{acknowledgements}

%
%
\bibliographystyle{aa}
\bibliography{draft}

\begin{appendix}\label{appendix:images}
\section{Fluxes of the subsample of EELGs with NIRSpec spectra}
\begin{table*}[!t]
    \centering
    \caption{Spectroscopic fluxes for the sample of EELG candidates with NIRSpec spectra ordered by increasing CEERS ID number. We include the MPT ID which is the identification code of the galaxies in the spectroscopic catalog in CEERS. We report fluxes from prism and medium-resolution grating. Fluxes are in cgs units of erg cm$^{-2}$ s$^{-1}$. In parenthesis, the uncertainties are reported.}
    \label{tab:nirspec}
\tiny{
\begin{tabular}{c|cccccccccc}
    ID &   MPT ID &  $z_{\rm spec}$ &     F([OII])$_{\rm prism}$ &     F([OII])$_{\rm MR}$ & F(H$\beta$)$_{\rm prism}$ &    F(H$\beta$)$_{\rm MR}$ &  F([OIII]$\lambda5007$)$_{\rm prism}$ &     F([OIII]$\lambda5007$)$_{\rm MR}$ &  F(H$\alpha$)$_{\rm prism}$ &     F(H$\alpha$)$_{\rm MR}$ \\
     &    &   &     10$^{-18}$cgs &     10$^{-18}$cgs & 10$^{-18}$cgs &    10$^{-18}$cgs &  10$^{-18}$cgs &     10$^{-18}$cgs &  10$^{-18}$cgs &    10$^{-18}$cgs \\\hline
  1253 &    44 &   7.10 &       -- &        -- &     <0.4 &       -- &  3.2(0.2) &        -- &        -- &        -- \\
  2149 &  3584 &   4.64 & 1.9(0.4) &        -- & 3.2(0.2) &       -- & 23.4(0.4) &        -- & 11.8(0.3) &        -- \\
  2166 &  1912 &   5.10 & 2.9(0.5) &        -- & 0.8(0.2) &       -- &  5.6(0.2) &        -- &  3.8(0.3) &        -- \\
  4176 &  1953 &   4.61 &       -- &        -- & 0.9(0.2) &       -- &  5.0(0.2) &        -- &  2.6(0.1) &        -- \\
  5040 &  3585 &   3.87 &       -- &  2.5(0.4) &     <1.7 & 2.0(0.2) &  1.1(0.3) & 20.9(0.5) &  0.6(0.2) & 10.5(0.3) \\
  8674 &  2355 &   6.12 & 0.8(0.2) &        -- &     <0.4 &       -- &  1.2(0.1) &        -- &  1.2(0.2) &        -- \\
  9290 &  3587 &   3.92 &       -- &        -- & 0.6(0.2) &       -- &  6.6(0.1) &        -- &  2.8(0.1) &        -- \\
 16056 &  2000 &   4.81 & 3.6(0.4) &        -- & 3.0(0.2) &       -- & 21.0(0.3) &        -- & 10.7(0.3) &        -- \\
 19984 &   323 &   5.67 &       -- &        -- & 1.6(0.2) & 2.2(0.3) &  8.2(0.2) &  7.5(2.3) &  5.4(0.2) &  6.1(0.2) \\
 21394 &   355 &   6.10 & 1.1(0.2) &      <4.3 & 1.3(0.1) & 1.1(0.3) &  7.9(0.2) &  7.7(0.2) &        -- &  3.5(0.2) \\
 25074 &   397 &   6.00 & 3.9(0.3) &  3.6(0.3) & 5.3(0.2) & 5.9(1.6) & 38.4(0.3) & 40.0(1.3) & 18.5(0.4) & 18.9(0.5) \\
 26436 &   428 &   6.10 &       -- &  0.7(0.2) &       -- &       -- &        -- &        -- &        -- &  2.9(0.3) \\
 27280 &   439 &   7.18 &       -- &        -- & 0.8(0.1) &       -- &  4.5(0.2) &        -- &        -- &        -- \\
 31338 &   498 &   7.18 &     <0.6 &        -- & 0.7(0.2) &       -- &  4.2(0.2) &        -- &        -- &        -- \\
 31339 &   499 &   7.17 &       -- &        -- &       -- &     <0.2 &        -- &  1.4(0.1) &        -- &        -- \\
 35306 & 82043 &   4.32 &       -- &        -- & 1.6(0.3) &       -- &  9.2(0.3) &        -- &  4.9(0.2) &        -- \\
 35645 & 82052 &   5.15 &       -- &        -- & 0.4(0.1) &       -- &  2.7(0.1) &        -- &  1.6(0.1) &        -- \\
 45809 & 80239 &   7.49 &       -- &        -- & 0.4(0.1) &       -- &  1.7(0.1) &        -- &        -- &        -- \\
 46186 & 82300 &   4.72 &       -- &        -- &     <0.5 &       -- &  1.6(0.2) &        -- &  0.8(0.1) &        -- \\
 46552 & 80244 &   7.01 &       -- &        -- & 0.3(0.1) &       -- &  1.1(0.1) &        -- &  <29506.8 &        -- \\
 48859 & 80671 &   5.74 &       -- &        -- &       -- &       -- &        -- &        -- &        -- &        -- \\
 53583 &   535 &   7.12 &       -- &        -- & 0.8(0.1) &       -- &  4.1(0.2) &        -- &        -- &        -- \\
 59817 &   792 &   6.26 &       -- &  2.5(0.3) &       -- & 0.8(0.2) &        -- &  6.5(0.3) &        -- &  2.3(0.3) \\
 59920 &  1027 &   7.82 &     <0.8 &      <0.7 & 1.6(0.2) & 1.2(0.2) &  9.8(0.3) &  9.5(0.5) &        -- &        -- \\
 61253 & 80710 &   6.55 &       -- &        -- &     <0.2 &       -- &  0.9(0.1) &        -- &  0.8(0.1) &        -- \\
 61419 &    24 &   9.00 &       -- &        -- &       -- &       -- &        -- &  2.2(0.3) &        -- &        -- \\
 70867 &  1236 &   4.48 & 1.2(0.4) &      <0.7 &       -- &     <5.9 &        -- &  2.3(0.2) &        -- &  1.3(0.2) \\
 78973 &  1305 &   4.28 &       -- &        -- &       -- & 0.5(0.1) &        -- &  2.5(0.2) &        -- &  1.3(0.1) \\
 79680 &  1038 &   7.19 &       -- &      <1.2 &       -- & 0.6(0.1) &        -- &  2.6(0.2) &        -- &        -- \\
 81061 &  1019 &   8.68 &       -- &  1.6(0.1) &       -- & 1.4(0.5) &        -- & 19.0(0.8) &        -- &        -- \\
 86030 & 80374 &   7.18 &       -- &        -- & 0.6(0.1) &       -- &  1.6(0.2) &        -- &        -- &        -- \\
 86830 & 80916 &   5.68 &     <0.8 &        -- & 1.1(0.1) &       -- &  7.5(0.2) &        -- &  4.4(0.2) &        -- \\
 87370 &  1374 &   5.00 & 4.4(0.4) & 11.6(0.5) & 2.7(0.2) & 9.8(0.5) & 20.9(0.3) & 65.8(0.9) &  9.9(0.2) & 31.4(3.5) \\
 90671 & 80083 &   8.64 &       -- &        -- &     <0.6 &       -- &  2.3(0.1) &        -- &        -- &        -- \\
 97883 & 83779 &   4.31 &     <0.9 &        -- & 0.9(0.2) &       -- &  8.9(0.2) &        -- &  4.8(0.2) &        -- \\
 98160 & 80432 &   7.48 &     <0.5 &        -- & 1.1(0.1) &       -- &  6.2(0.2) &        -- &        -- &        -- \\
100152 & 81063 &   6.09 &       -- &        -- & 0.7(0.1) &       -- &  7.4(0.2) &        -- &  3.2(0.2) &        -- \\
100312 & 83856 &   4.56 &       -- &        -- & 1.5(0.1) &       -- &  4.3(0.1) &        -- &  4.8(0.2) &        -- \\
100621 & 81068 &   6.27 &       -- &        -- & 0.7(0.1) &       -- &  4.4(0.2) &        -- &  2.1(0.1) &        -- \\\hline
\end{tabular}}
\end{table*}
\end{appendix}
\end{document}